\documentclass[12pt,aps,nofootinbib,superscriptaddress]{revtex4-2}
\usepackage[utf8]{inputenc}
\usepackage{amsfonts}
\usepackage{amssymb}
\usepackage{amsmath}
\usepackage{amsthm}
\usepackage{mathtools}
\usepackage{mathrsfs}
\usepackage{graphicx}
\usepackage{tabularx}
\usepackage{enumerate}
\usepackage{siunitx}

\usepackage[margin=1.1in]{geometry}
\newcommand{\eq}[1]{Eq.~\eqref{eq:#1}}

\newcommand{\eqss}[3]{Eqs.~\eqref{eq:#1}, \eqref{eq:#2}, and \eqref{eq:#3}}
\renewcommand{\sec}[1]{Sec.~\ref{sec:#1}}

\newcommand{\appx}[1]{Appendix \ref{appx:#1}}
\newcommand{\fig}[1]{Fig.~\ref{fig:#1}}

\newcommand{\tab}[1]{Table~\ref{tab:#1}}

\newcommand{\be}{\begin{equation}}
\newcommand{\ee}{\end{equation}}

\begin{document}
\title{Laser-Wakefield-Driven Photonuclear and Laser-Driven DD Fusion Neutron Sources for Fast Neutron Capture: A Start-to-End Simulation Study}
% Comparing Laser-Wakefield-Driven and Laser-driven DD Fusion Neutron Sources for Fast Neutron Capture: A Start-to-End Simulation Study
% Laser-Driven Neutron Sources for Fast Neutron Capture and Applications: start-to-end simulations

\author{Ou Z. Labun}
\affiliation{Center for High Energy Density Science, Department of Physics, University of Texas at Austin, Austin, Texas, USA}
\author{D. D. Phan}
\affiliation{Center for High Energy Density Science, Department of Physics, University of Texas at Austin, Austin, Texas, USA}
\author{L. Labun}
\affiliation{Department of Physics, University of Texas at Austin, Austin, Texas, USA}
\affiliation{Tau Systems Inc, Austin, Texas, USA}
\author{M. L. Klebonas}
\affiliation{Center for High Energy Density Science, Department of Physics, University of Texas at Austin, Austin, Texas, USA}
\author{Calin Hojbota}
\affiliation{Center for High Energy Density Science, Department of Physics, University of Texas at Austin, Austin, Texas, USA}
\author{Philip Franke}
\affiliation{Tau Systems Inc, Austin, Texas, USA}
\author{Sam Yoffe}
\affiliation{SUPA Department of Physics, University of Strathclyde, Glasgow, Scotland, G4 0NG, UK}
\author{Rahul Kumar}
\affiliation{Tau Systems Inc, Austin, Texas, USA}
\author{B. M. Hegelich}
\affiliation{Center for High Energy Density Science, Department of Physics, University of Texas at Austin, Austin, Texas, USA}

\date{March 2026}

\maketitle

\section{Introduction}
A wide range of scientific and technological applications depend on the ability to extract an accurate neutron energy spectrum from a pulsed source. The common methodology is time-of-flight (TOF) measurement, in which a short burst of neutrons is produced at a known time, and the energy of each neutron is inferred from its travel time to a detector at a known distance \cite{Gus07, And16}. The energy resolution achievable by TOF scales inversely with pulse duration, which is, up to the Heisenburg quantum uncertainty limit, the shorter the neutron burst, the finer the energy discrimination at a given flight path. Peak flux determines the statistical quality of the measurement within that burst. Ultra-short, high-flux neutron pulses therefore unlock a broad class of TOF-based measurements that are inaccessible to conventional longer-pulse sources, and high-intensity ultrashort pulse lasers have emerged as a uniquely capable platform for producing them.

The applications enabled by this principle span multiple fields:

{\bf Nuclear structure and cross-section measurements.} Fast neutron resonance spectroscopy (FNRS) uses TOF to resolve isotope-specific resonances in neutron interaction cross sections, providing direct access to nuclear level structure and reaction rates \cite{Gue13, Gus07,Moo23}. Single-shot resonance spectroscopy has been demonstrated with laser-driven sources \cite{Yog23}, and prospects for ultrahigh energy resolution using laser wakefield acceleration (LWFA)-driven photonuclear sources at 100 Hz have been outlined \cite{Wan23}.

{\bf Non-destructive material analysis and isotope identification.} Neutron resonance transmission analysis (NRTA) exploits isotope-specific TOF absorption dips to determine elemental and isotopic composition non-destructively inside bulk objects with applications in materials science, cultural heritage, and nuclear safeguards \cite{Zim22, Koi24, Pos09}. Laser-driven sources are particularly attractive here because their small source size enables compact, point-like geometries \cite{Mim22}.

{\bf Dynamic compression and shock thermometry.} In high-energy-density experiments, neutron resonance Doppler broadening measured via TOF provides a non-contact thermometry diagnostic for shocked materials on nanosecond timescales, which is a measurement impossible with conventional sources \cite{Yua05,Lan24}. The picosecond pulse duration of laser-driven sources is naturally matched to the timescales of dynamic compression experiments.

{\bf Nuclear security and active interrogation.} Short, intense neutron pulses enable time-gated detection of fission signatures from special nuclear materials, with demonstrated single-pulse interrogation capability using laser-driven sources \cite{Rot17, LANL24}.

{\bf Nuclear astrophysics.} Direct measurement of neutron capture cross sections on neutron-rich nuclides relevant to heavy element nucleosynthesis requires instantaneous fluxes far exceeding conventional sources \cite{Che19, Jia23}.

Across these applications, the key figures of merit are the same: how short is the neutron pulse, and how many neutrons does it deliver? Conventional accelerator-based sources, such as spallation sources and reactor-based facilities, have pulse durations typically limited to nanoseconds or longer, with peak fluxes constrained by their large source volumes \cite{And16}. High-intensity ultrashort pulse lasers, operating at peak powers from terawatts to petawatts, drive neutron production through laser-plasma interactions and offer a qualitatively different regime: pulse durations naturally on the order of picoseconds to tens of picoseconds, source sizes of order (10s \si{\micro\meter})$^3$, and peak fluxes that have recently surpassed $>10^{22}$/cm$^2$/s \cite{Jia23}, which is more than two orders of magnitude beyond the best spallation sources. These properties make laser-driven sources a compelling and rapidly developing platform across the range of TOF-based neutron applications described above.

While laser-driven neutron production encompasses several methods, three account for the majority of the experimental literature: ``pitcher-catcher'' configuration laser-accelerated ion beams, deuterium-deuterium (DD) bulk fusion, and laser wakefield acceleration (LWFA)-driven photonuclear.  A fourth, cluster fusion \cite{Dit99}, offered an early entry to laser-driven neutron production, but does not offer any advantages for neutron ToF applications.  Among the applications these three enable, r-process cross-section measurement is the most demanding in terms of both flux and pulse duration, and will be the primary design interest in this paper. The r-process is responsible for synthesizing approximately half of all elements heavier than iron \cite{Rem06,Cow21}, yet neutron capture cross sections on the relevant neutron-rich nuclides remain largely unmeasured, as the current indirect methods cannot access r-process kinematics directly, leaving model predictions uncertain by factors of 2–10 \cite{Rau00, Mum16}. A complete simulation-based assessment of whether laser-driven sources can meet these demands has not previously been reported; prior work \cite{Hil21} examined source feasibility without incorporating a full chain from laser-plasma interaction through neutron transport to capture kinematics. This paper provides that assessment, using DD bulk fusion and LWFA photonuclear as candidate sources evaluated against this benchmark.

The choice of DD bulk fusion and LWFA photonuclear neutron production for in-depth comparison reflects three considerations.  Ion acceleration mechanisms such as target normal sheath acceleration (TNSA) often occur simultaneously with the bulk fusion inside the laser-heated plasma, making the two contributions inseparable without careful diagnostics. This makes isolation of the bulk mechanism a diagnostic goal in itself. Second, the mechanisms generate neutron pulse of widely differing duration: bulk fusion and LWFA photonuclear produce orders of magnitude shorter duration neutron pulses. For TOF applications, this is not a minor difference but a qualitative distinction in achievable energy resolution. Third, DD bulk fusion exhibits an approximately isotropic angular distribution, in contrast to the forward-peaked TNSA and LWFA sources. Isotropy simplifies diagnostic characterization and enables simultaneous multi-angle TOF measurements, making it the practical choice for developing and validating neutron spectroscopy platforms.  The physical and diagnostic basis for this choice is elaborated in Section II.

This paper presents a complete start-to-end simulation comparison of DD bulk fusion and LWFA photonuclear neutron sources, evaluated specifically for their suitability as ultra-short, high-flux sources for fast neutron resonance spectroscopy.  We describe in detail the technical requirements and physics extracted from the complete simulation chains for both source types (particle-in-cell for laser-plasma interaction, Monte Carlo particle transport, and neutron capture Monte Carlo), compared for laser parameters including both $\sim 1$ J/30 TW (UT3-class tabletop) to $\sim$250 J/ PW (PHELIX/ELI-NP-class).  We derive and compare scaling laws governing neutron yield in pulse lasers, the neutron pulse duration, and peak flux for each source type, and compare to long-pulse results where applicable.
Scaling laws combined with quantitative examples allow quantitative comparison of neutron pulse characteristics, especially, duration, spectrum, angular distribution, and peak flux, as figures of merit for TOF-based applications.  We conclude with a discussion of  high-level advantages and disadvantages of each source type and where each is preferred.  We intend this to guide experiment design given the recent changes in facility availability, e.g. the TPW going offline (see \appx{TPW}) and new petawatt (PW)-class facilities coming online via the European Extreme Light Infrastructure (ELI).

% We did not do: A diagnostic framework for separating DD bulk fusion contributions from TNSA pitcher-catcher contributions in mixed experimental geometries

The paper is organized as follows. Section II describes the physical mechanisms and theoretical framework. Section III presents the simulation methodology. Section IV collates and compares results to discuss pros and cons of different neutron sources for applications. Section V concludes.

\section{Physics of laser-driven neutron sources}

Experimental work spanning 1999–2025 has demonstrated ultra short pulse laser-driven neutron production across a wide range of laser systems, target configurations, and laser pulse parameters.  We tabulate reported results in Appendix \ref{sec:lit_survey}. For TOF-based applications, the relevant characterization goes beyond neutron yield to include pulse duration $\tau_n$, source size, and angular distribution $\Delta\Omega$, which are the properties that determine energy resolution and geometric flexibility in TOF measurements.  These quantities come together in the peak flux, roughly estimated as
\begin{align}
    F\simeq \frac{N_n}{\tau_n\Delta\Omega}
\end{align}
and many of the novel applications targeted by laser-driven neutron sources call for maximizing the peak flux $F$ while retaining a large overall neutron number $N_n$ in the pulse.  Both are important because higher flux increases time and energy resolution and potentially allows access to novel nuclear physics via multiple-neutron events, while the neutron number controls the total signal generated and hence signal to noise.  The source size can limit both the peak the flux and the geometry of the sample/target to be irradiated by the neutrons.  However,
these three quantities, pulse duration $\tau_n$, source size, and angular size $\Delta\Omega$, are rarely reported together in the present literature.  Appendix \ref{sec:lit_survey} summarizes available data separated into tables according to the three physical mechanisms/methods of neutron production: pitcher-catcher configurations for ion-driven reactions, fusion reactions in laser-driven plasmas, or laser wakefield accelerator-driven photonuclear reactions.  As we will see, the We briefly review the physics of each mechanism with the purpose of developing scaling and efficiency estimates for neutron production.

\subsection{Pitcher-catcher ion-driven}

A laser-irradiated primary target accelerates ions via target normal sheath acceleration (TNSA).  As the lightest neutron-containing ions, deuterons are strongly favored for neutron production purposes, though proton-driven reactions have been considered as well \cite{petrov2012generation}. The deuterons impinge on a secondary converter target, producing neutrons via stripping, break-up and light fusion reactions, all of which are more probable for converters composed of light nuclei for maximizing yield.  Lithium and berrylium in particular offer multiple channels for MeV neutron production, e.g. the $^7$Li(d,n)$^8$Be reaction advantageous for its relatively high energy release $Q\simeq 15$ MeV.  With multi-MeV deuteron energies from TNSA, the neutron source can be forward-peaked but is generally hemispherical in angular distribution, with yields scaling strongly with laser energy at PW-class facilities \cite{Rot13, Kle18}.  Among the three major types of neutron sources discussed in this paper, TNSA produces the highest neutron yield for equivalent laser energy \cite{Mir17}. This is because TNSA is an efficient acceleration mechanism to transfer laser energy into a large number of light ions, such as deuterons.  

We use TNSA as a historical and contextual baseline rather than a primary comparison objective.  For the goal of this paper, the neutron beam characteristics are too limited by the geometric setup, i.e. the pitcher-catcher distance and thickness of the converter. The neutron energy spectrum is typically broader compared to other neutron sources. Fusion reactions can add MeV kinetic energy to the neutron (as in the case of $^7$Li(d,n)$^8$Be) but break-up and stripping reactions are often endothermic.  Accounting for all channels in the converter, a broad range of outgoing neutron energies are possible. Combining the quasi-exponential distribution of TNSA-accelerated deuterons with the kinematics of the various production channels and thick target effects broadens the spectrum even further so that the neutron energies can extend from much less than an MeV to several 10s of MeV.  The neutron pulse duration is set by the non-relativistic ion flight time to the converter,
\begin{align}
     \tau_n\simeq L_{\rm p-c}/\Delta v_i
\end{align}
Depending on the geometric setup i.e. the pitcher-catcher distance and the thickness of the converter in the individual experiment, the pulse of MeV neutrons is typically nanoseconds long, arising from in-flight dispersion. The neutron source size is millimeter to centimeter scale, arising from the volume covered by the ion beam on the converter.  Even in cases of forward-beaming, the beam divergence is large, half-angle $\gtrsim 20$ degrees, inherited from the ion beam  $\gtrsim 20$ degree divergence \cite{Wil01, Schollmeier:2009kca} and increased by scattering kinematics \cite{Rot16, Mac13, Fer09}.

To predict the neutron yield and efficiency, we need models for the conversion of laser energy into ion  kinetic energy and the dependence of neutron yield on ion energy.  Scaling laws and yields estimates are synthesized in the appendix.  The result is
\begin{align}
    Y_n\propto a_0\ln(a_0)\,g\!\left(\frac{E_\mathrm{th}\ln a_0}{m_ea_0}\right)
\end{align}
where $g(x) \to \mathrm{const}$ for $x \ll 1$ (reactions well above threshold, $T_p \gg E_\mathrm{th}$) and $g(x) \sim e^{-x}$ for $x \gtrsim 1$ (reactions near or below threshold).

\subsection{Bulk fusion} 
To create a fusion-based neutron source, a laser directly irradiates a dense deuterated target, heating and compressing the plasma.  Within the bulk of the plasma, deuterons undergo fusion reactions producing 2.45 MeV neutrons in the center-of-mass frame.  This mechanism produces a quasi-monoenergetic neutron spectrum centered at 2.45 MeV and an approximately isotropic angular distribution, which is unique among laser-driven sources in the TW-to-PW range considered here at TOF-relevant flux levels. Recent results at the Texas Petawatt Laser (TPW) obtained a peak flux exceeding $10^{22}\si{\per\centi\meter\squared\per\second}$ \cite{Jia23}, making this the current benchmark for laser-driven neutron flux. 

\begin{figure}
    \centering
    \includegraphics[width=0.8\textwidth]{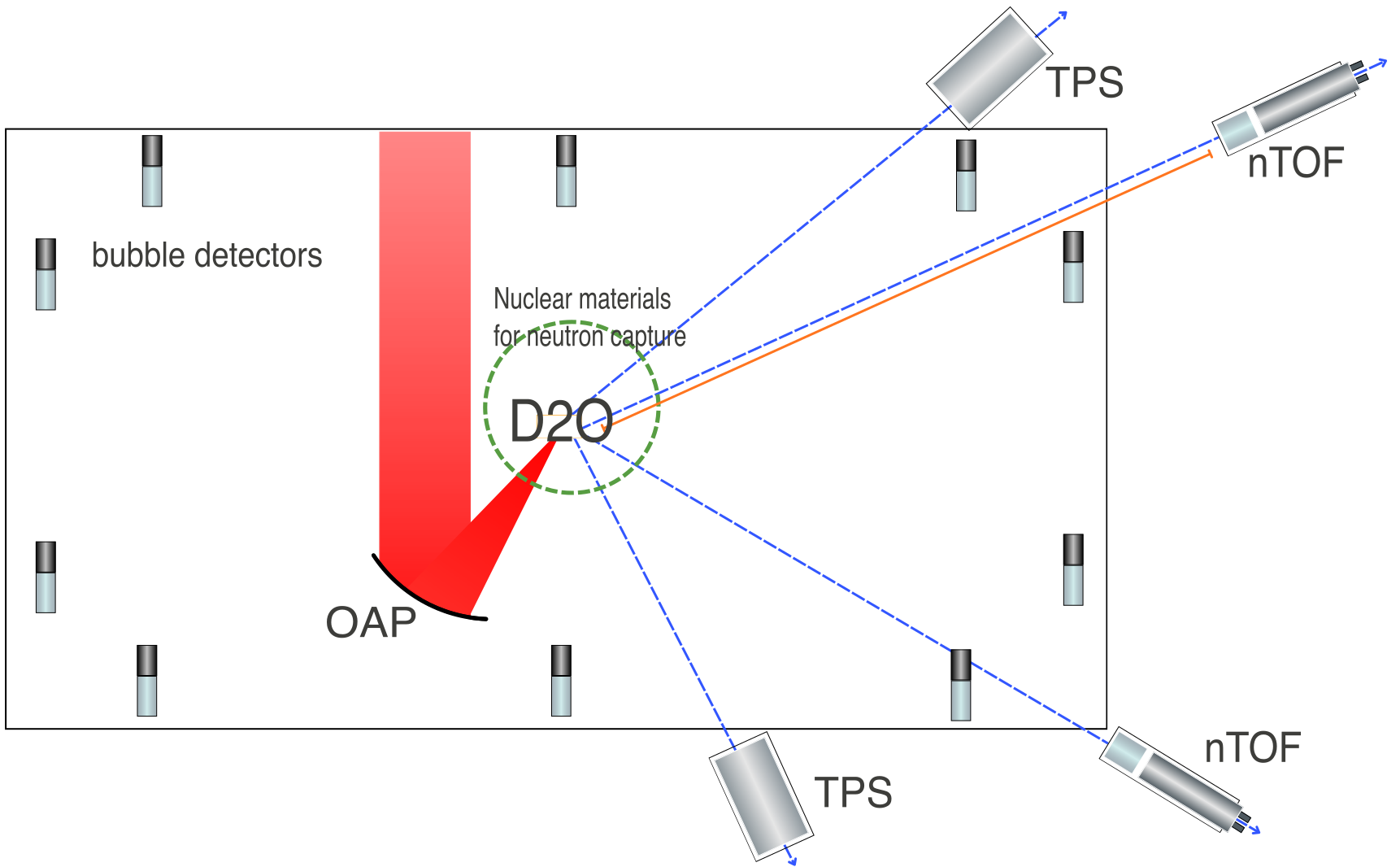}
    \caption{\label{fig:DD_layout}  An example layout of the final focusing chamber in a PW-class laser system for a DD fusion-based neutron source, showing the schematic geometry of the laser, plasma and nuclear waiting material. }
\end{figure}

Deuterium is chosen being relatively common and readily available both in pure and compound forms, and of neutron-producing reactions, the DD fusion reaction, $D + D \to n + {}^3\mathrm{He} 3.27$ MeV, has the second lowest barrier with a peak cross section around 400 keV CoM energy.  While the deuterium-tritium reaction becomes rapid at even lower ion energy, tritium as a reactant presents logistical challenges.  

DD bulk fusion generally achieves the smallest source size and narrowest energy spread but produces the lowest neutron yields of the three.  The fundamental reason is that DD fusion requires $\gtrsim 10$ keV ion energies while maintaining sufficient density and confinement time, all of which are challenging with short-pulse lasers and reflect basic compromises in the physics.

The source size is determined by the laser-heated plasma volume, which is several times the laser spot size in length scale 10–100 \si{\micro\meter}.  Coupling laser energy into the plasma is enhanced by achieving relativistic transparency, allowing the laser pulse to ``bore'' into the target and deposit energy beyond the front surface skin depth.  Ion acceleration experiments have proven the high efficiency of this mode of laser-ion coupling.  However, because the goal here is to increase the ion kinetic energy while still in a high-density region of the target, the target should be designed and prepared to enhance hole-boring (also described as ``pistoning'') dynamics.  (We wish to avoid the radiation pressure and the blow-out afterburner regimes where ions are accelerated ``as a block'' or by the laser breaking through and driving a thin cloud expanding from the rear surface.)  Achieving hole-boring accelerates ions in two ways: (1) ion reflection from the compressed electron layer at front of the moving hole/piston can result in quasi-monoenergetic ion spectra, as predicted by PIC simulations under ideal conditions, and (2) dissipation of the strong charge separation fields created by the channeling, which may occur through collisionless shock heating, instabilities, and decompression heating.  Due to the geometric plus kinematic scaling of the fusion yields, the latter population of quasi-thermal or multi-Maxwellian ion yields orders of magnitude more neutrons, and the direct irradiation yield models of Ref \cite{Lab23} apply.

Intuitively the laser energy coupling into the plasma is enhanced by a density gradient on the front surface, enabling the laser to self-focus and pass the critical density surface by inducing relativistic transparency more gradually.  This intuition agrees with PIC simulations \cite{Lue18,Lue20} and experiments \cite{Hor21}.  

The neutron pulse duration is dominated by the plasma disassembly time:
\begin{align}
    \tau_n\simeq \tau_\mathrm{exp}=\frac{R_0}{c_s},
\end{align}
where $\tau_\mathrm{exp}$ is the plasma expansion time, $R_0$ is the length scale of the heated volume (proportional to the laser focal spot size) and $c_s=\sqrt{T_e/m_i}$ is the ion sound speed. For the laser focal spot 5-20 \si{\micro\meter}, $\tau_\mathrm{exp}$ ranges from 0.5 to tens of picoseconds for electron temperatures from 100 keV to 10 MeV.  This ensures the neutron pulse duration is at least one order of magnitude smaller than the TNSA neutron pulse duration \cite{Zwe00} but significantly contributes to limiting the neutron yield.

The optimal laser pulse thus balances the hydrodynamic expansion timescale against fusion timescale. Longer laser pulses deposit more total energy and sustain the plasma at fusion temperatures longer, producing more neutrons, while too long a pulse $\tau_\ell\gtrsim \tau_\mathrm{exp}$ allows the plasma to expand and cool before fusion is complete \cite{Pre98, Mor98, Gib05, Mul10, Dai12}.  The optimal laser pulse length for the solid target for neutron production is thus a few picoseconds, provided the intensity is high enough to achieve volumetric heating.  %Experiments with ultra-short pulse lasers ($\tau_\ell<1\,\si{\pico\second}$) help study the transition between beam-target and thermonuclear regimes \cite{Bos92}, which requires the non-thermalized regime accessible only with short pulses, Inertial Confinement Fusion (ICF) Core Diagnostics, as well as applications requiring ps-duration multi angle neutron bursts for time-resolved diagnostics \cite{Bry73,Bal90,Mur14}. 

The neutron energy spectrum is quasi-monoenergetic and peaked near 2.45 MeV in the lab frame, with a narrow thermal width set by the keV-range deuteron temperature, which is far narrower than the quasi-exponential TNSA distribution. 

Considering that volumetric ion heating dominates the yield, the estimate and scaling is considerably simpler than TSNA.  We need only how the effective deuteron temperature scales with laser intensity $T_D(a_0)$ in this regime.  Then the neutron yield is estimated from the typical thermonuclear expression,
\begin{align}
    Y_n = \frac{1}{2} n_D^2 \langle \sigma v \rangle_{DD} V_\mathrm{hot} \tau_{\text{exp}},
\end{align}
where $n_D$ is the deuteron density, $\langle \sigma v \rangle_{DD}(T_D)$ is the thermally averaged cross section, and $V_\mathrm{hot}\propto w_0^2L_\mathrm{HB}$.  This offers a good estimate, even if the deuteron energy distribution has supra-thermal tail or second, hotter Maxwellian distribution in the $>100$ keV range (as seen in our simulations): in this case, the relevant $T_D$ is the mean energy of the hotter population, which covers more of the peak DD fusion cross section.  Higher laser energy at fixed intensity means a larger focal volume, expanding the fusion-active region. Higher intensity at fixed spot size raises the plasma temperature, increasing the fusion rate per unit volume but not necessarily the source size.  

Optimizing the bulk fusion yield requires preparing a target with low enough electron density (ideally a few times $n_c$), possibly utilizing pre-plasma to support the transition to hole-boring, and balancing sufficient laser intensity for hole-boring against increasing the heated volume.  The benefits are clear though.  The neutron pulse duration is the shortest, the neutron energy spread narrowest and the source size smallest.  These features favor the bulk fusion source for applications needing energy precision or a point-like source, e.g., high-resolution neutron imaging that highly depends on the source geometry \cite{Nor98}.

\subsection{Laser wakefield accelerator-driven photonuclear}\label{sec:LWFAscaling}
A laser operating in the wakefield acceleration regime in an underdense plasma accelerates electrons to a few tens of MeV to multi-GeV energies \cite{Taj79}. These electrons generate bremsstrahlung photons in a converter, which in turn drive photonuclear ($\gamma$,$n$) reactions via the giant dipole resonance. The resulting neutron source is forward-peaked and collimated, with a broad, harder energy spectrum extending well above 2.45 MeV. This mechanism is accessible at TW-class tabletop laser systems operating at high repetition rates, with demonstrated yields of $1.6\times 10^5$ n/sr at TW-class tabletop systems \cite{Jia17} and $3.1\times 10^7$ n/sr at higher-energy facilities \cite{Val25}.

\begin{figure}
    \centering
    \includegraphics[width=0.8\textwidth]{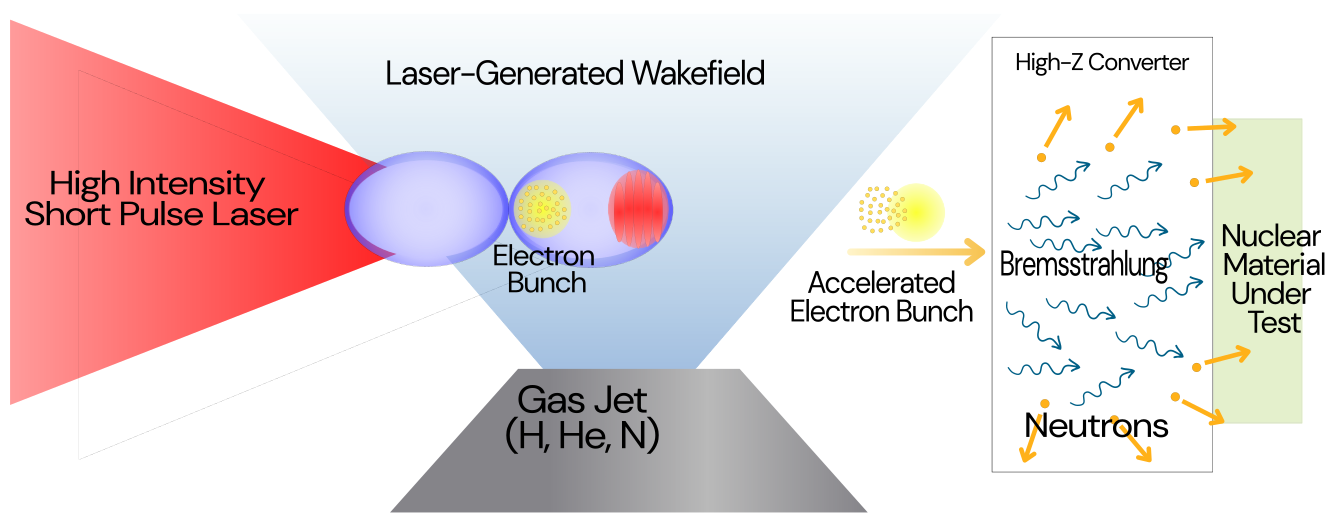}
    \caption{\label{fig:LWFA_layout}  Schematic of an LWFA photoneutron source.}
\end{figure}

The photonuclear cross section peaks at the Giant Dipole Resonance around 10 MeV to 25 MeV depending on target nucleus but has a long tail. Photonuclear reactions produce neutrons with a broad energy spectrum mirroring the bremsstrahlung spectrum above threshold around 2 MeV. For a tungsten converter and GeV \cite{Lee06} electrons, neutron energies span from $\sim 1$ MeV to $\sim 100$ MeV, but the majority of neutrons tend to be in the few-MeV range \cite{Li22, Fen20}. The typical width of the neutron spectrum is larger than the spectral width from DD bulk fusion but narrower than TNSA, with a much longer high-energy tail.

Neutron beam divergence is the key advantage of LWFA. Because the electrons are relativistic, the bremsstrahlung photons and resulting photoneutrons can be strongly forward-boosted by relativistic beaming. The photon emission cone half-angle scales roughly as $\sim 1/\gamma$, which transfers to the neutrons for those produced well above the GDR energy.  In this way, LWFA can produce a more collimated beam of high energy $E_n\gtrsim 10$ MeV neutrons than TNSA, which makes it useful for radiography or directed applications.  At lower electron beam energies $E_e\lesssim 100$ MeV, the lower neutron energy ($E_n<10$ MeV) results in a more isotropic distribution, which can be somewhat affected by neutron re-scattering in the converter

The source size is dominated by the bremsstrahlung photon scattering in the converter and the electron stopping range.  By the time the electron beam reaches the converter target, its $\lesssim 10$ mrad divergence causes it to expand to 0.5-5 mm, depending on distance.  The MeV-photon mean free path and radiation length of tungsten, a typical converter material, are a few millimeters.  Convolving the electron beam divergence and the photon opening angle, we estimate most neutrons are produced within 1 cm of the beam axis.  Neutrons are produced until the electrons are stopped inside or pass out of the converter.  The effective source size is consequently typically a few millimeters in diameter by several centimeters long.  For low-divergence, high-energy neutrons, the millimeter source size and 10 mrad divergence are good for radiography.

The neutron pulse duration is determined by the electron transit time in the converter $\tau_n\simeq L/c$, where $L$ is the converter thickness and the upper limit is set by electrons remaining relativistic while emitting over the entire converter length. For a general thick target $L\gtrsim L_\mathrm{stop}$, $\tau_n$ is around tens of picoseconds \cite{Yog23}.

To maximize photoneutron yield, we should maximize the total LWFA energy efficiency and achieve the highest total electron beam energy $E_b=\int E_e (dN/dE)dE_e$.  The reason is that bremsstrahlung photon yield increases with both electron energy $E_e$ and beam charge, as long as the endpoint of the spectrum exceeds the GDR region.  Using photoneutron scaling laws from the literature \cite{Mao96},
\begin{align}
\frac{N_n}{N_e}& \simeq Y_\mathrm{thick}(Z)\ E_e[\mathrm{MeV}]   \\
 Y_\mathrm{thick}(Z)&\simeq 8\times 10^{-6}\left(Z^{1/2}+0.12\,Z^{3/2}-0.001\,Z^{5/2}\right)\frac{\mathrm{n}}{\rm MeV}
\end{align}
where the thick target yield models depend on atomic number $13\leq Z\leq 82$ of the converter.  A simpler analytic approximation is $Y_\mathrm{thick}\simeq  2\times 10^{-5}Z^{0.67}$, which may be substituted with slightly less accuracy near the ends of the $Z$ range \cite{Swa79,Mao96}.  Since the total beam energy in quasi-monoenergetic beam is $E_b\simeq N_eE_e$, we have
\begin{align}
    N_n&\simeq \kappa (E_e)\,Y_\mathrm{thick}(Z)\,  \frac{E_b[J]}{e}\\
     &\simeq 1.5\times 10^{6}\,\left(Z^{1/2}+0.12\,Z^{3/2}-0.001\,Z^{5/2}\right)\,\kappa(E_e)\left(\frac{E_\ell}{J}\right)^{\!1.09}
\end{align}
where the factor $\kappa(E_e)$ accounts for the multiplicity of photons in the GDR region and switches between $\simeq 1$ for $E_e\ll 200$ MeV and $2-3$ for $E_e\gg 200$ MeV.  On the first line, the converter physics is separated in the first two terms and the electron beam source physics is captured by the beam energy. In the second line, we have substituted the 95\% upper prediction interval performance for $E_b$ extracted from the literature by Ref \cite{Lab25}.  A more conservative estimate from the average past performance $E_b\simeq (3\,\mathrm{mJ})(E_\ell/\mathrm{J})^{0.9}$ reduces the leading coefficient by a factor 10.  Optimization then depends on what laser class is used.  

Widely-available $\sim 1$ J lasers easily produce electron beams with $>100$ MeV centroid energy, meaning that much of the bremsstrahlung spectrum is not used efficiently, being above the GDR region.  The LWFA should be therefore designed to produce as high charge beams as possible at more moderate energy, for example by using ionization injection and/or above-optimal plasma density.  Electron beam energy spread has no negative impact on the neutron source performance, so this common side effect of high beam loading is not important.  With the assumption that total beam $E_b$ can be traded 1-1 between electron energy and beam charge, a highly simplified exponential model of the bremsstrahlung spectrum suggests that tuning the electron beam energy around $\simeq 100$ MeV maximizes photon yield in the GDR region and hence neutron yield \cite{Lab23a}.  We estimate that with up to 100 pC beam charge, $10^8$-$10^9$ neutrons per shot should be achievable, hence up to $10^{10}$-$10^{11}$ n/s at 100 Hz repetition rate.

Larger PW-class lasers can easily produce multi-GeV electron beams.  De-tuning them to produce high charge $\sim 100$ MeV would not be an effective use, considering their more limited repetition rate.  At multi-GeV energies, the bremsstrahlung radiation develops into a shower, with cascading photon energy substantially increasing the photon number in the GDR range, given a thick enough converter ($\gtrsim 5X_0$).  In this case, nanocoulomb-scale beam charges at several GeV centroid energy can yield $10^{10}$-$10^{11}$ neutrons per shot, with a significant number ($10^5$-$10^6$) high energy and strongly beamed.  At $>1$ GeV electron energies, direct electron-nucleus reactions also become probable, producing neutrons via fission and/or spallation.  However the cross section and yield are so low that the resulting high energy neutrons $E_n\gtrsim 500$ MeV are probably not useful.  Repetition rate and injection method now recommend the lowest laser energy that can achieve multi-GeV beams (eg 20 J) with nanoparticle injection \cite{Ani24} to boost beam charge.  The 1-1 energy-rep rate trade off possible with Tm:YLF lasers cite{} could enable neutron sources with flux average flux comparable with spallation sources \emph{including} a more collimated high-energy component.

Higher laser energy allows operation at lower plasma density with the increased plasma wavelength p , which increases the plasma bubble size and therefore the electron beam source size. However, lower density also means less divergence from the wakefield due to weaker trapping potential inside the plasma bubble. Higher laser energy also accelerates electrons to higher energies, meaning a higher Lorentz factor $\gamma$ [Gil05, Sar14]. All these together lead to the more collimated neutron beam with the narrower coned forward-boosted bremsstrahlung photons [Esa09]. It is also the reason LWFA is the most collimated of the three sources.
Pulse length must satisfy a highly effective operating condition: the pulse duration should roughly match the plasma wave half-period: laserp2c . Too long a pulse drives plasma waves inefficiently, leading to electrons dephase early with lower final energy and eventually worse neutron collimation. Too short a pulse causes insufficient energy coupling into the wakefield and degrades $\gamma$ and therefore the 1/$\gamma$ beaming angle[Gun22, Pom14]. Matched pulse length is critical for efficient LWFA and for achieving maximum electron energy. In fact, pulse length is the most important parameter for LWFA collimation of the three sources. 

The two sources compared in this paper, DD bulk fusion and LWFA photonuclear, represent the two ends of the laser-driven neutron source design space most relevant for TOF-based applications: one optimized for extreme per-shot flux and spectral purity at PW-class facilities, the other for high repetition rate and broad spectral reach at TW-class tabletop systems and PW-class facilities. The following sections develop the theoretical framework and simulation methodology needed to compare them quantitatively on TOF performance grounds.

\section{Simulation Methodology}

%The experimental survey in Section II establishes that DD bulk fusion and LWFA-driven photonuclear sources span overlapping laser energy ranges yet exhibit qualitatively different neutron pulse properties. 
To move from the qualitative, scaling comparison of the previous section to quantitative one, we require simulations that connect laser parameters to neutron source figures of merit across the full 1J to 250J range.  Our specific goal is to determine which source delivers superior neutron pulse performance at a given laser facility.  %This section builds the simulation.

\begin{figure}
\centering
\includegraphics[width=0.6\textwidth]{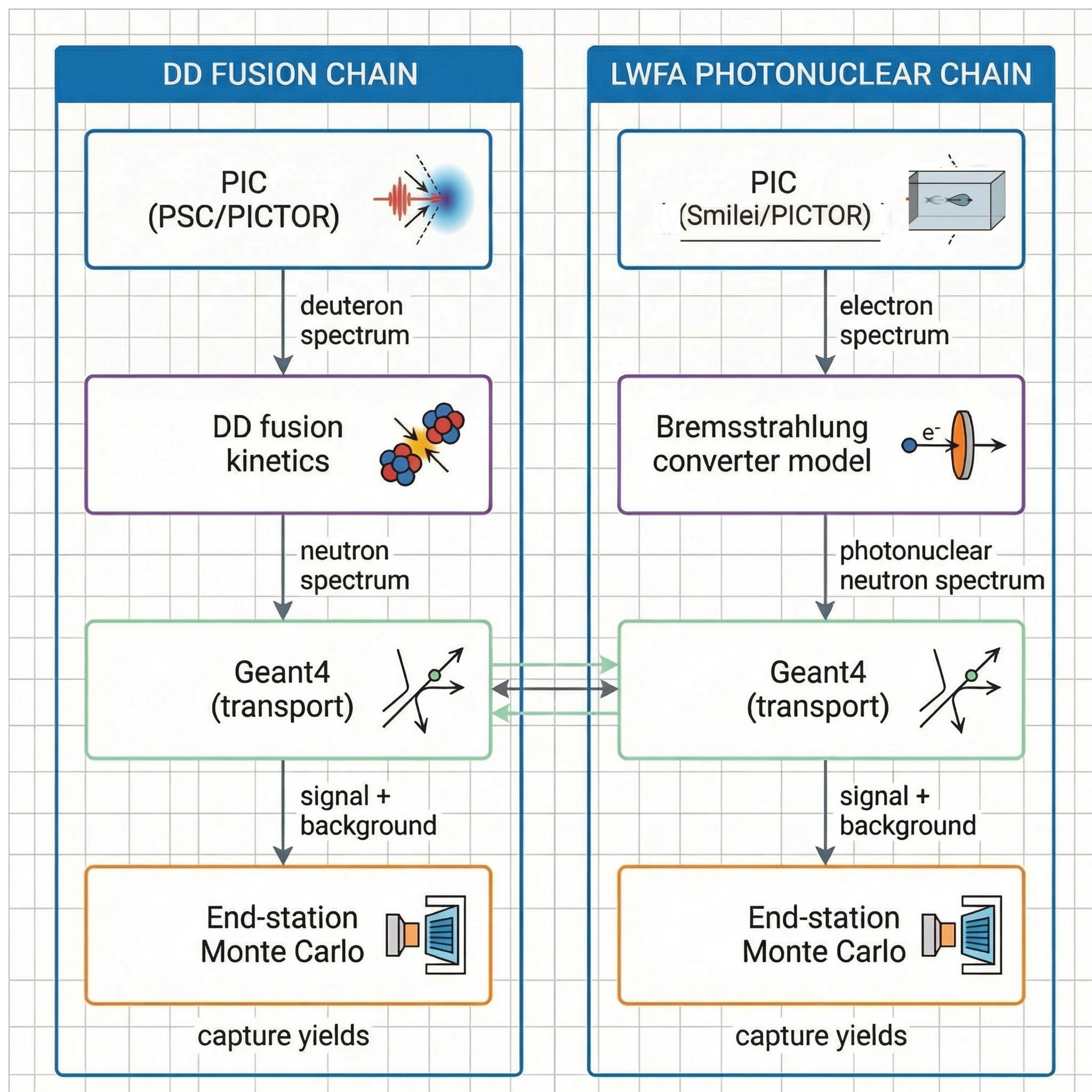}
    \caption{\label{fig:simulation_workflow} Simulation workflows and example codes for bulk fusion and LWFA-driven photonuclear neutron sources.}
\end{figure}

Simulating a short-pulse laser neutron production experiment from first principles requires three distinct stages: the kinetic laser-plasma interaction handled by a Particle-in-Cell (PIC) code, the particle transport through geometry and encoded nuclear interaction handled by GEANT4 Monte Carlo simulation, and the signal vs. background detection for particle source application handled case by case at the end-station with an event generator. Figure \ref{fig:simulation_workflow} summarizes the complete chain for both source types in parallel. Each subsection below describes the method, highlights the most important physics, discusses the limitations and presents the results produced from an example case at that stage.

\begin{table}
\centering
    \begin{tabular}{l l c c c c}
    Facility & Location & $\lambda_\ell$ [\si{\micro\meter}] & $E$ [J] & $\tau$ [fs] & Peak $a_0$  \\ \hline
    PHELIX & GSI, Darmstadt, DE & 1.053 & 200 & 500 & 28  \\
    ELI-NP & Magurele, RO & 0.8 & 25-250 & 25 & 41-129 \\
    Jupiter & Livermore, CA, USA & 1.054 & 300 & 700 & 8 \\
    ZEUS & East Lansing, MI, USA & 0.8 & 75 & 25 & 153 \\ \hline
    \end{tabular}
    \caption{\label{tab:laser_facilities} Laser facilities and nominal laser parameters }
\end{table}

\begin{table}[h]
\centering
    \begin{tabular}{lcccl}
    \hline
    Material & $n_e$ [cm$^{-3}$] & $n_e / n_{\mathrm{cr}}$ ($\lambda_\ell=800$\,nm) & Unit cost & Availability \\
    \hline
    D$_2$   & $6\times10^{22}$   & 29  & \$35/L    & Commercial, immediate \\
    CD$_2$  & $3.6\times10^{23}$ & 214 & \$1150/g  & Commercial; low-cost academic\\[-3mm]
    & & & & not ready until late 2024 \\
    D$_2$O  & $3.3\times10^{23}$ & 197 & \$1300/L  & Commercial, immediate \\
    CD$_4$  & $\sim1.6\times10^{23}$ & 94 & \$290/L & Commercial, immediate \\
    \hline
    \end{tabular}
    \caption{\label{tab:deuteron_source_mats} Deuteron source material properties and availability}
\end{table}

\subsection{Stage 1: Particle-in-cell simulations}

The first stage simulates the particle acceleration during the laser-plasma interaction. When the focused pulse reaches the target, it drives relativistic kinetic dynamics on femto- to picosecond timescales over micron-scale spatial domains. Whether the target is a surface of near-solid density or a gas distribution of lower density, this part of the interaction is handled by a Particle-in-Cell (PIC) code, which solves macro-particle trajectories self-consistently with electromagnetic fields on a spatial grid. An advantage of PIC is that it predicts the six-dimensional phase space distribution of the accelerated particles.  The disadvantage is high computational cost.  For this reason, identifying the most important observables and potentially developing computationally efficient surrogate models for those observables would have high value for designing and optimizing neutron sources.
%deuterons and electrons for the DD fusion chain, electrons for the LWFA chain, 

\subsubsection{Bulk fusion}

The challenge for PIC simulations of bulk fusion is two-fold. First, accurately capturing the laser-solid density interaction requires small spatial and temporal steps to resolve the Debye length $\lambda_D=\left(\epsilon_0T_e/e^2n_e\right)^{1/2}$ and the relativistic motion of electrons $\Delta t\leq \lambda_\ell/ca_0$.  These resolution requirements make the simulations computationally expensive, so that typically at most a handful of the most realistic three-dimensional simulations are run.  

The second challenge is that the fusion reaction yield must be computed from the simulation information.  The most intensive approach is to add by hand a collision term to the particle-in-cell algorithm.  Considering that even near-barn cross sections for light nuclear reactions such as $d(d,n)^3\mathrm{He}$ imply a $\lesssim 10^{-3}$ fraction of deuterons react, it is assumed that adding such collisions to the Boltzmann-Vlasov system solved by PIC is a good-enough approximation.  By adding the reactions to the PIC operations in each timestep, the in-situ ion distribution functions can be used to compute the yield.  Since so few such reactions occur, the loss of the reactants and the creation of the products is usually neglected.  For this same reason, saving the complete ion distribution at regular intervals throughout the simulation and computing the reaction yield offline offers an estimate of similar accuracy.  The simplest, but fastest method is to evaluate a reduced expression for the yield using only the ion energy distribution.  As explained above, in practice, two kinematic regimes dominate the yield: collisions within the slower ($E\lesssim 1$ MeV) but more isotropic deuteron population, and collisions between fast ($E\gtrsim 1$ MeV) deuterons, accelerated by charge separation fields, and slower deuterons in the bulk.

We use the PSC \cite{Ger16} and PICTOR \cite{Kum17} codes in 2D geometry to model the laser interaction with a solid or liquid deuterated target. Table \ref{tab:laser_facilities} lists the parameters of the laser facilities considered in this study. Among these, PHELIX most closely matches the TPW conditions under which the benchmark neutron flux result \cite{Jia23} was obtained. ELI-NP represents a realistic case given the recent advances in Ti:sapphire systems offering higher energy and intensity.  It also has a well-equipped nuclear diagnostics laboratory. The target is a deuterated solid slab with an exponential density gradient extending from the front surface.  The pre-plasma is modeled by quasi-neutral self-similar expansion \cite{Mor79,Mor03} (with no continuous heating \cite{Sai25}) according to the TPW temporal contrast profile, which has an amplified spontaneous emission threshold $\sim 14$ ps before peak.  the choice of target material is informed by Table \ref{tab:deuteron_source_mats}, which compares candidate deuterated materials. CD$_2$ and D$_2$O are both viable in terms of electron density, but given the price inflation of CD$_2$, we adopt D$_2$O as the baseline target for this study.  

\begin{figure}
\centering
    \includegraphics[width=0.48\textwidth]{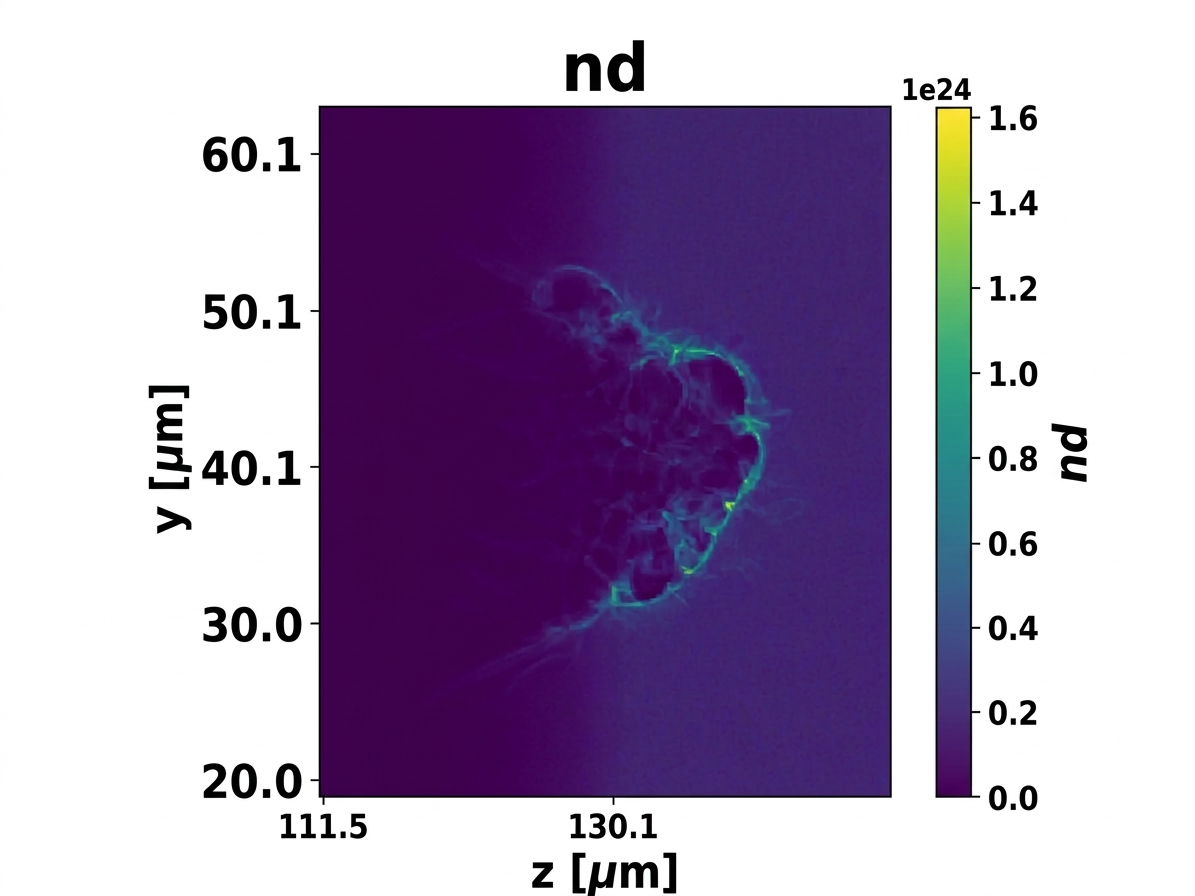}
    \includegraphics[width=0.48\textwidth]{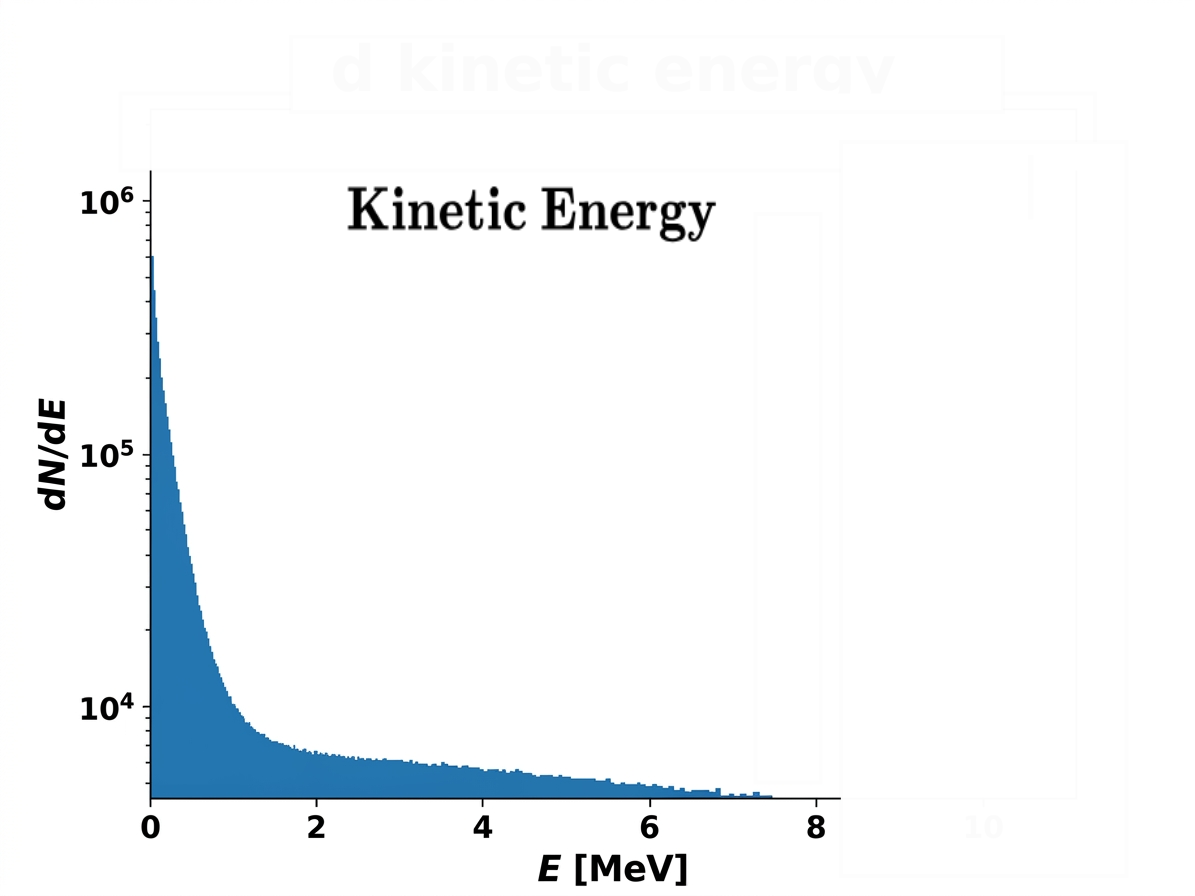}
    \caption{\label{fig:PHELIX_bulk} 2D PIC simulation of the laser-plasma interaction at approximate PHELIX conditions.  Left: deuteron density distribution shortly after peak of laser pulse arrives.  Laser propogates from the left to the right.  Right: deuteron spectrum at the same time.}
\end{figure}

We run 2D PIC simulations, being 2-3 orders of magnitude faster to complete than 3D, cognizant of the systematic errors in modeling the laser-plasma interaction at high-intensity and solid density \cite{Sta17}.  To convert 2D PIC neutron yields to 3D absolute numbers, we apply a simple geometric scaling. In a 2D Cartesian simulation, the out-of-plane dimension is effectively infinite, so particle numbers are naturally expressed per unit length (units of m$^{-1}$). The physical 3D neutron yield is recovered by multiplying by the effective out-of-plane extent, taken to be the laser focal spot radius $w_0$:
\begin{align}
    N_n^{(3D)}\simeq N_n^{(2D)}w_0
\end{align}
where $N_n^{(2D)}$ is the neutron yield per unit length obtained by integrating the DD fusion reaction rate over the 2D simulation domain,
\begin{align}
N_n^{(2D)} =\Delta t \int  n_D^2(x,z) \langle\sigma_{DD} v\rangle dA \,.
\end{align}
Here $n_D$ is the deuteron number density from the PIC simulation, $\langle \sigma_{DD} v\rangle$ is the velocity-averaged DD fusion cross section evaluated at the local deuteron temperature, $dA$ is the 2D area element, and $\Delta t$ is the interaction time over which fusion occurs (e.g. $\tau_{exp}$) or a relevant simulation time interval over which the data-based estimate applies. The assumption of cylindrical symmetry is well-motivated for a linearly polarized laser pulse focused to a near-circular spot on a planar target, and introduces an estimated uncertainty of a factor of a few in the absolute yield.

An example of laser heating and deuteron acceleration in the plasma is shown in Figure \ref{fig:PHELIX_bulk}. The deuteron density distribution and energy spectrum  are shown shortly after the laser peak. The simulations predict a deuteron kinetic energy spectrum that is approximately exponential at low energies ($E_d \lesssim 1$ MeV) with a temperature of a few hundred keV, ensuring a significant population in the peak region of the DD fusion cross section near $E_d \sim$ 1 MeV. Approximately 5\% of laser energy is transferred to deuterons within a focal-spot scale volume, from which we estimate a neutron yield $\sim 10^9$, and a peak neutron flux exceeding $10^{22}$/cm$^2$/s at the source and $>10^{20}$/cm$^2$/s at approximately 0.5 cm from the source.  Under these conditions, an expectation for the broadening of the neutron spectrum can be obtained using the semi-analytic formulas of Ref \cite{App11}.  The peak neutron energy and thermal broadening both increase linearly with deuteron temperature for $T_D<Q_{DD}=2.45$ MeV and increase more rapidly $T_D\gtrsim Q_{DD}$.  The neutron line also develops a high energy tail breaking the symmetry of the spectrum around the peak.  The Brysk formulae based on saddle-point approximation \cite{Bry73} significantly overestimate the width and cannot capture the line asymmetry.

The simulation also provides important information for diagnostics design and background characterization.  Only $\sim 10^{-3}$ of deuterons undergo fusion; the remaining unfused deuterons propagate out of the plasma and contribute the dominant background, motivating the Pb shielding design in Stage 2. Because the contributing deuterons have kinetic energy similar to the energy released in the reaction, the resulting neutron momentum distribution is approximately isotropic. The plasma additionally expands against the laser propagation direction, which constrains detector placement to the sides and rear of the target rather than along the forward laser axis. 

In summary, the important outcomes of the PIC simulation include not only more quantitative predictions of the signal (neutron yield $N_n$ and spectrum $dN_n/dE_n$) but also quantitative information about the plasma environment and background eg $dN_i/dEd\Omega$.  Considering that laser-plasma interactions in the high-intensity, high-density regime are sensitive to initial laser and plasma conditions that cannot be exhaustively explored in simulation, we use the results as a model of known (in)accuracy, informing design decisions and interpretation, rather than a definitive account.

\subsubsection{Laser wakefield acceleration}
We use Smilei, PICTOR, and FBPIC to simulate an underdense gas jet target at UT3 TW-class and TPW PW-class laser conditions. The parameters for both conditions are listed in \tab{lwfa_params}. Like other PIC codes used for laser-plasma accelerator studies, PICTOR implements 2D and 3D Cartesian problem spaces with moving windows that track the laser through its interaction with the plasma \cite{Kum17}. Unlike other PIC codes, PICTOR supports explicit 3D cylindrical problem spaces, reducing the computational cost of 3D simulations by one to two orders of magnitude compared to 3D Cartesian simulations and providing greater azimuthal resolution without the parallelization penalty of Fourier transforms required by spectral algorithms \cite{Kum24}. Because the LWFA bubble has approximate azimuthal symmetry about the laser axis, 3D cylindrical geometry is both physically appropriate and computationally efficient for these simulations.

\begin{table}[h]
\centering
\begin{tabular}{lcc}
\toprule
\textbf{Parameter} & \textbf{UT3} & \textbf{TPW} \\
\hline
Laser energy (J)            & 0.2            & 53 \\
Pulse duration (fs)         & 35             & 135 \\
Spot size ($\mu$m FWHM)     & 3.5            & 55 \\
Wavelength ($\mu$m)         & 0.8            & 1.054 \\
Peak $a_0$                  & 4              & 3.1 \\
Electron density (cm$^{-3}$)& $2.0 \times 10^{19}$ & $6.0 \times 10^{17}$ \\
\hline
\end{tabular}
\caption{LWFA PIC simulation parameters for UT3 and TPW laser conditions.  Laser energies are corrected to the energy contained in the central spot, since the simulations model only a single gaussian focal spot. }
\label{tab:lwfa_params}
\end{table}

The two simulation conditions span qualitatively different plasma regimes. At UT3 conditions ($n_e = 2\times 10^{19}$ cm$^{-3}$), the plasma wavelength is $\lambda_p\simeq 7$ \si{\micro\meter}, comparable to the laser spot size of 3.5 \si{\micro\meter}, placing the interaction near the matched bubble regime. At TPW conditions ($n_e = 6\times 10^{17}$ cm$^{-3}$), the plasma wavelength increases to $\lambda_p\simeq 43$ \si{\micro\meter}, the bubble is substantially larger, and electrons are accelerated over a longer dephasing length to higher final energies. Both simulations are near the matched pulse condition $\tau_{\ell}\simeq \lambda_p/c$ \cite{Esa09}, maximizing energy coupling into the wakefield and therefore electron beam charge and energy.

\begin{figure}
    \includegraphics[height=4.5cm ]{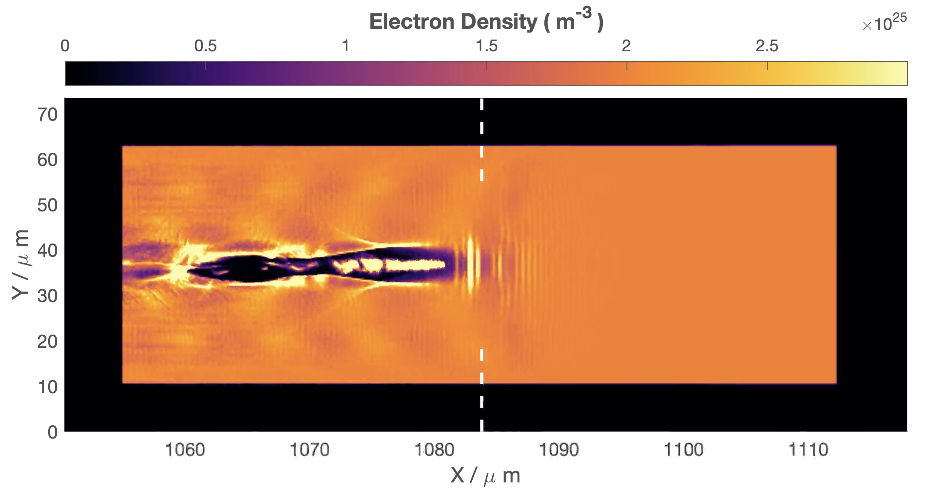}
    \includegraphics[height=4.5cm ]{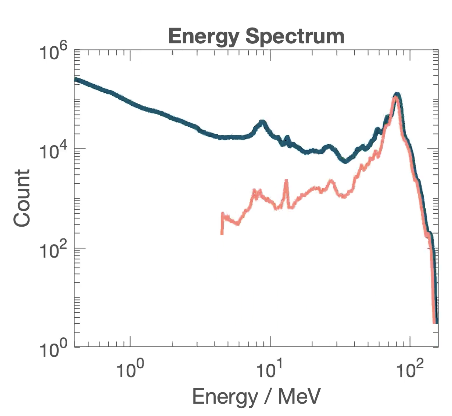}
    \caption{\label{fig:UT3_LWFA} 3D simulation of UT3 LWFA using PICTOR.  Left: electron density in a slice passing through laser axis after $\sim 1$ mm propagation.  Right: electron spectrum showing a broad peak around $\sim 90$ MeV, in agreement with recent performance.  The pink line includes only electrons within 10 mrad of the axis.}
\end{figure}
 
The injection mechanism differs between the two conditions. At UT3 conditions, the lower pulse energy suggests using higher plasma density and ionization injection, in which electrons from a high-Z dopant gas are ionized directly inside the bubble and trapped earlier than self-injection. At TPW conditions, the lower plasma density and larger bubble allow self-injection, in which background plasma electrons are trapped at the back of the bubble once the wakefield steepens sufficiently. The two mechanisms produce electron beams with different energy spreads and charges, which propagate through to different bremsstrahlung spectra and neutron yields in Stage 2.

\begin{figure}
    \includegraphics[width=0.8\textwidth]{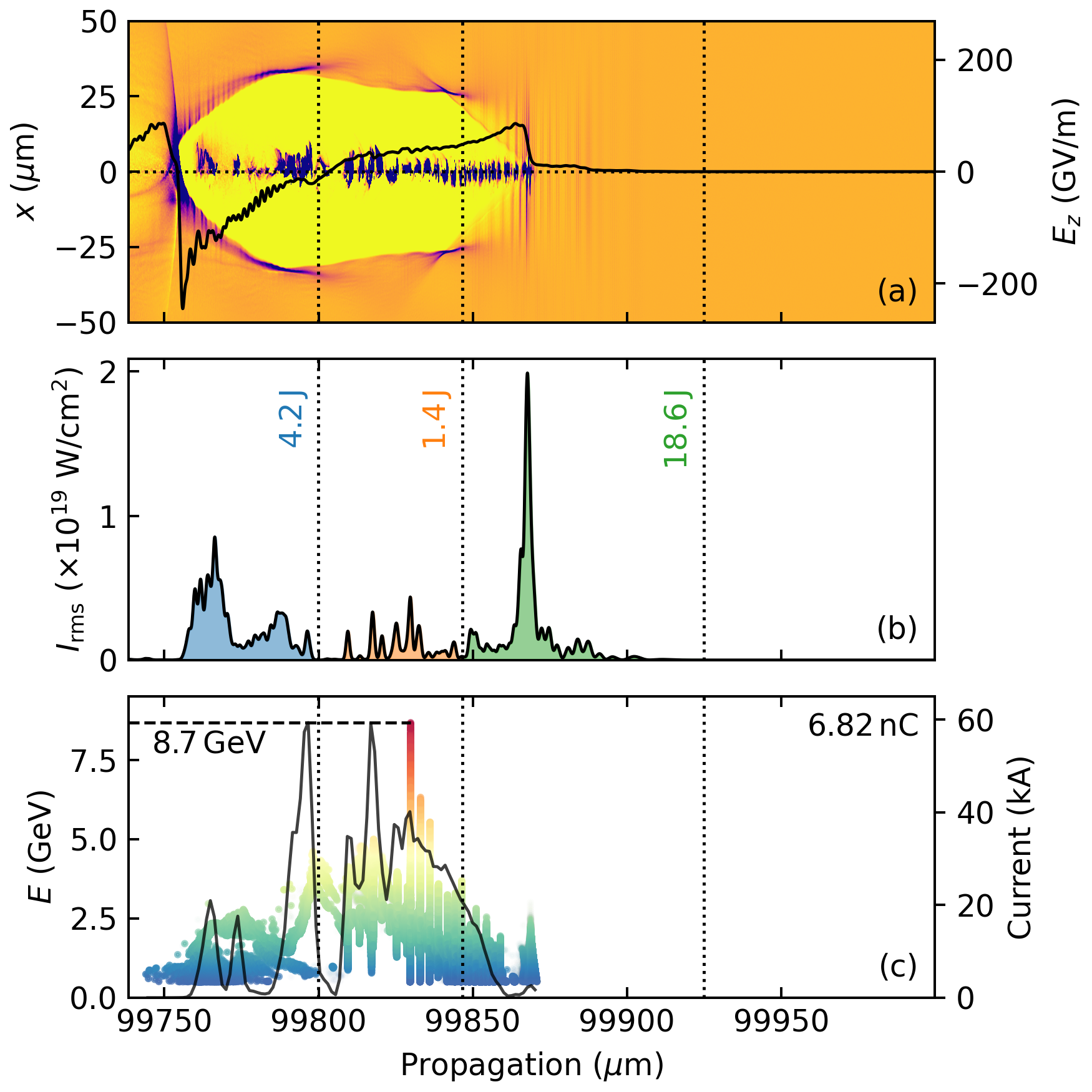}
    \caption{\label{fig:TPW_LWFA} 3D (azimuthal mode decomposed) simulation of TPW LWFA.  Top: electron density in a slice passing through laser axis after $\sim 10$ cm propagation.  Middle: Laser intensity longitudinal distribution.  Bottom: Longitudinal $x-vt,E$ phase space}
\end{figure}

An example of the LWFA beam generated at UT3 conditions is shown in Fig. \ref{fig:UT3_LWFA}; the corresponding TPW simulation is shown in Fig. \ref{fig:TPW_LWFA}.  The simulations produce beams in good agreement with observed performance in the lab.  Recent UT3 performance has been 95 MeV  centroid energy with 25 pC of beam charge \cite{Fra24}. However the TPW has met expectations of $>2$ GeV, nanocoulomb beams only with the addition of nanoparticles \cite{Ani24}.

The primary outputs used as inputs to Stage 2 are the electron energy spectrum and beam charge, because, as described in \sec{LWFAscaling}, neutron source performance depends primarily on the electron beam energy and charge.  As long as the beam is quasi-monoenergetic (as opposed to say Maxwellian), the precise energy spread has less impact, being convolved into the typical bremsstrahlung spectrum and then photoneutron cross section dependence.  Since only beam energy and charge are important, a (quasi-)three-dimensional simulations are necessary, but existing reduced-physics simulations and surrogate models can be used effectively: numerical artifacts, for example from using only one azimuthal mode or failing to suppress numerical Cherenkov, have little effect on the final neutron source predictions as long as they do not impact injection.

\subsection{Stage 2: Neutron transport and background modeling}

The second stage handles particle transport through the experimental geometry. For the DD fusion chain, keV to MeV-temperature deuterons confined within the 10–100 \si{\micro\meter} interaction region undergo thermonuclear fusion, producing neutrons that subsequently travel centimeters to meters through the target chamber. For the LWFA chain, the electron beam is transported to a high-Z converter, where bremsstrahlung photons drive photonuclear reactions and produce neutrons. In both cases, we simulate this stage with Geant4, a Monte Carlo code that encodes nuclear interaction cross sections and probabilistic particle transport through material geometries.  MCNP is the primary alternative code for this use.  The inputs are particle phase space distributions, either directly extracted from Stage 1 simulations or modeled to match the important features.  The outputs are the neutron energy spectra and angular distributions at the waiting targets, along with the photon and charged-particle backgrounds.

\subsubsection{Bulk fusion shielding, chamber, signal and background}
The laser-plasma interaction produces a high number of fast charged particles in addition to the desired neutrons.  In addition to transport of the neutrons to the waiting targets, we conduct both (1) a shielding study that determines what material is needed to stop the high-energy deuterons from the laser-driven plasma, and (2) a yield study to estimate radioisotope production at the waiting targets from residual background neutrons and photons. The shielding materials themselves produce tertiary neutrons and gammas via nuclear reactions that constitute an additional background. The rapid neutron capture events are handled separately in Stage 3; they are not modeled within Geant4, as no accurate GEANT4 physics model for this process is currently available.  A schematic overview of the geometry of the laser-driven plasma neutron source and waiting material was provided in \fig{DD_layout}.

To ground our discussion in concrete case, the shielding is designed for PHELIX PW-class laser conditions with a D$_2$O target, as used in the DD fusion PIC simulations in Stage 1. Only $\sim 10^{-3}$ of deuterons undergo fusion before escaping and producing neutrons; the vast majority of deuterons propagate in the forward and transverse directions relative to the laser axis, escape at the rear surface of the D$_2$O target, and must be blocked before reaching the nuclear waiting targets.  To calculate shielding requirements, we consider two cases: (1) the hard component of the escaped deuteron spectrum from the Stage 1 PIC simulations, modeled as a Maxwell-Boltzmann distribution with mean $\langle E_d\rangle = 2.5$ MeV and extending up to $\sim$30 MeV, and (2) the deuteron spectrum measured experimentally on the TPW \cite{Jia23}.  If the faster deuterons are stopped, then the soft component of the deuteron spectrum seen in the PIC simulations will also be stopped and at $\lesssim 1$ MeV energy does not contribute  photons or neutrons than can penetrate.  We include deuterons up to 30 MeV because the higher ion energy opens more reaction channels and penetrates a greater distance of shielding material, and even a few such background events would significantly contaminate the neutron capture signal.  To account for possible residual target effects, we add 1 mm of D$_2$O in front of the Pb-shield.  Geant4 shows that 4 mm of Pb stops all deuterons up to 30 MeV while allowing sufficient neutrons to pass. 

\begin{figure}
\centering
\includegraphics[width=0.43\textwidth]{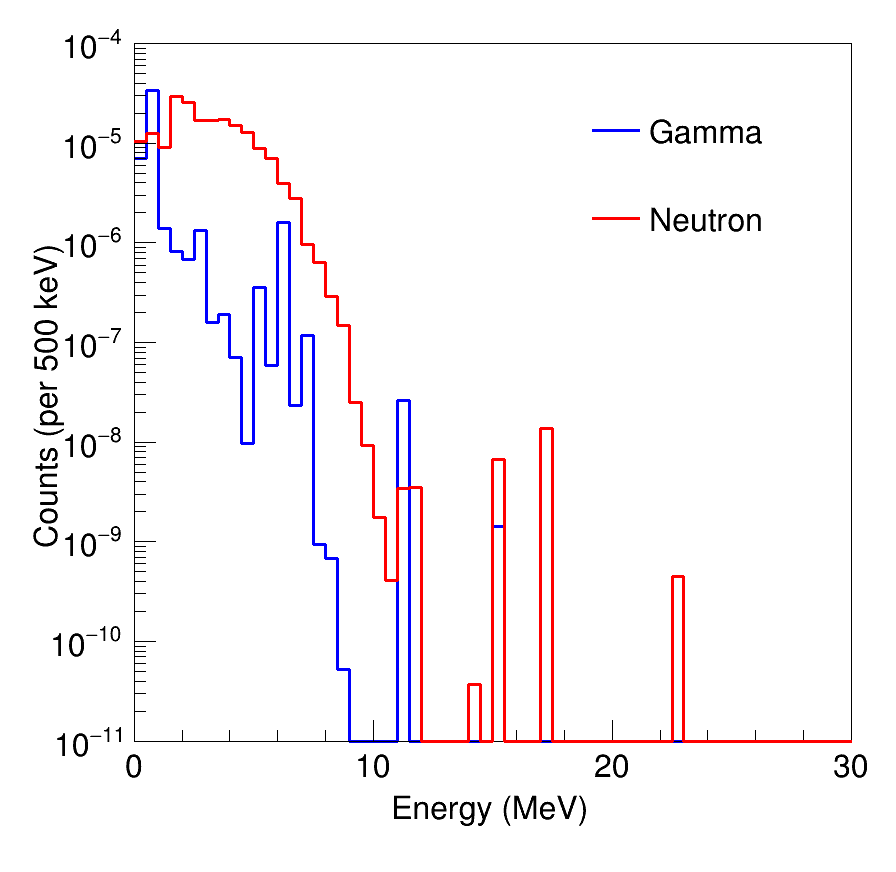}
\includegraphics[width=0.43\textwidth]{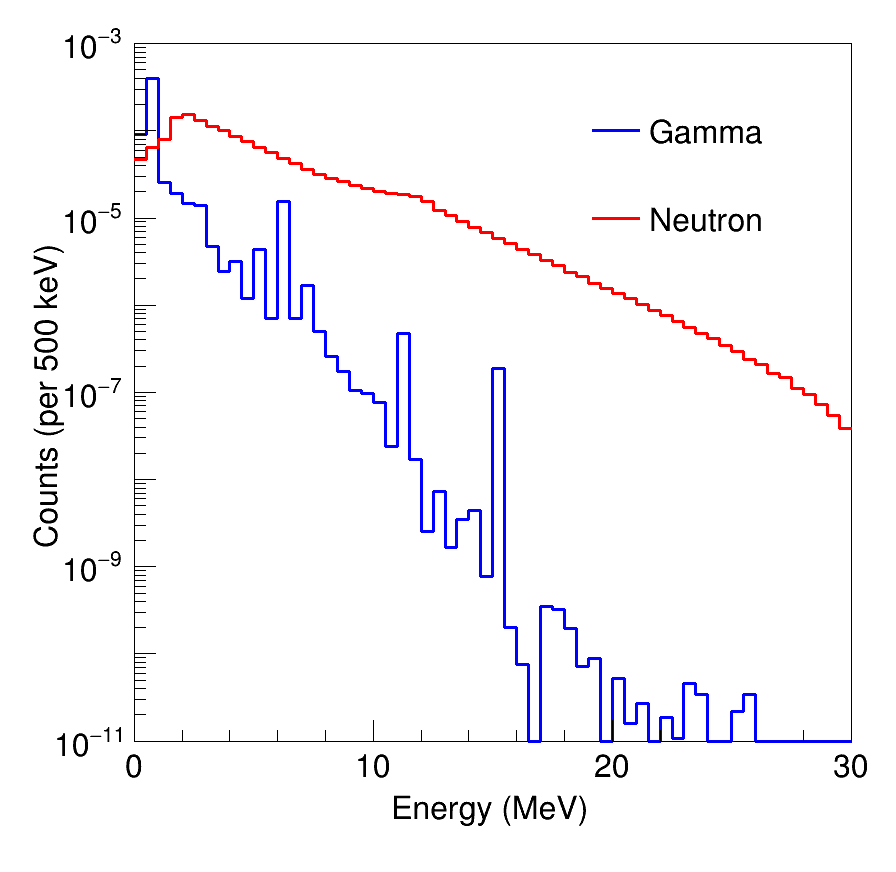}
\includegraphics[width=0.43\textwidth]{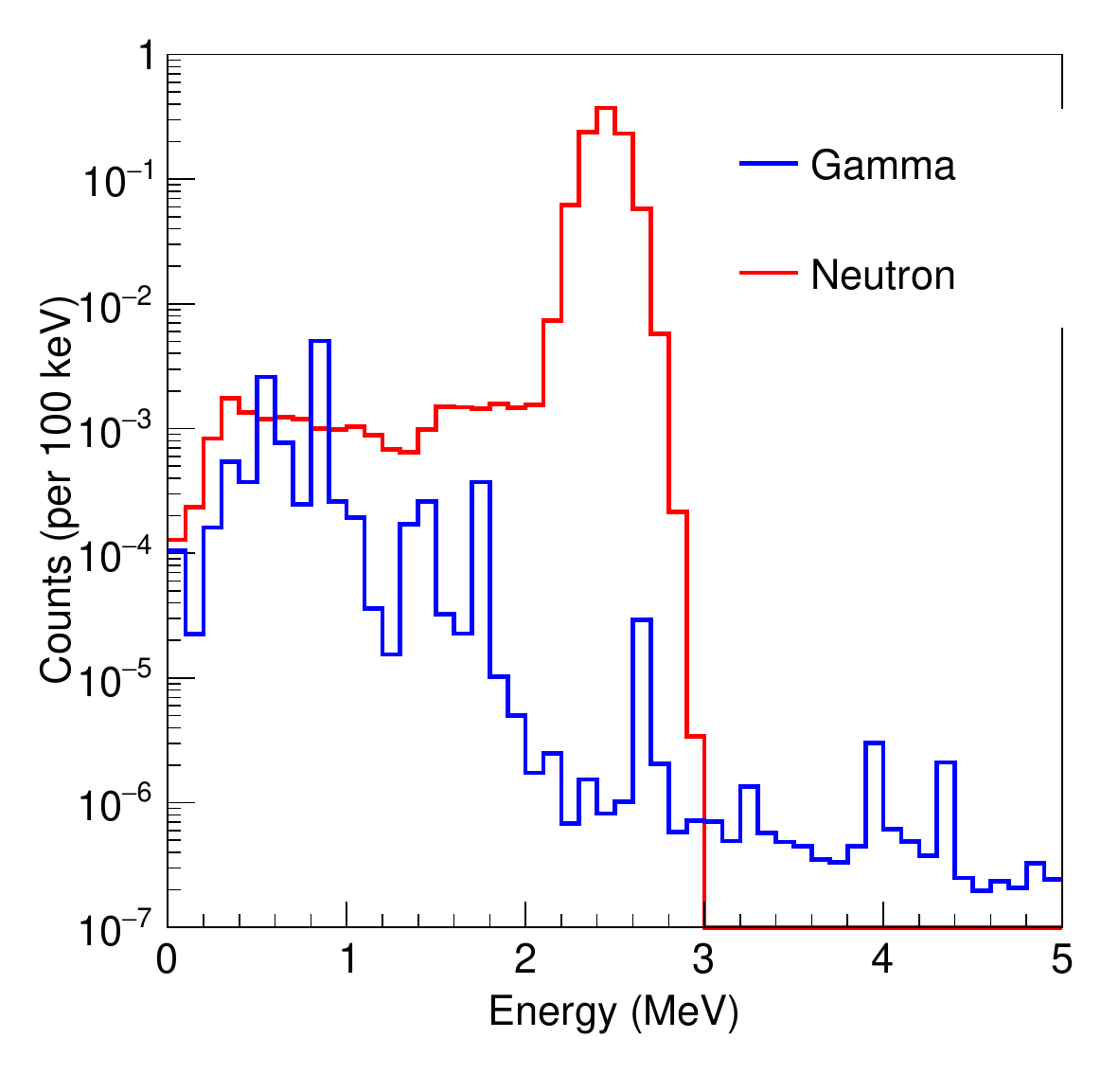}
    \caption{\label{fig:GEANT4_shielding} GEANT4 simulated particle spectra at the nuclear waiting targets after transport through 1mm  D$_2$O + 4mm Pb shielding.  All deuterons up to 30 MeV have been stopped.  Upper left: spectrum of secondary neutron and photons reaching the waiting target given the deuteron input spectrum from PIC simulation.  Upper right: spectrum of secondary neutron and photons given the deuteron input spectrum from Ref \cite{Jia23}.  Lower: spectrum of neutrons and secondary photons due to the signal distribution of near-2.5 MeV neutrons passing through the shielding.  The bin width and hence vertical scale differs between left, middle and right}
\end{figure}

The background neutron and photon spectra at the nuclear waiting targets, induced by deuteron interactions with the D$_2$O residual and Pb shielding, are shown in \fig{GEANT4_shielding} left (Model 1, PIC-derived Maxwellian input) and middle  (Model 2, experimental input). In both cases the background is dominated by neutrons. The PIC-derived Maxwellian spectrum (Model 1) yields substantially lower background than the experimentally derived spectrum (Model 2), approximately $8\times$ fewer neutrons and $13\times$ fewers photons. The difference reflects the harder high-energy tail of the experimental spectrum, which places more deuterons above nuclear reaction thresholds in Pb and D$_2$O.

The signal neutron spectrum is modeled as a Gaussian centered at 2.45 MeV with $\sigma_E=0.1$ MeV.  As discussed above, this true fusion spectrum will be asymmetric and could have a larger width, considering the likely deuteron distribution.  By erring on the low side for width, this model gives a conservative estimate of the sensitivity of the delivered neutron spectrum to the shielding.  The signal spectra at the nuclear waiting targets, induced by fusion neutron interactions with the D$_2$O and Pb shielding, are shown in \fig{GEANT4_shielding} right. The 2.45 MeV peak is largely preserved after transport through the shielding, confirming that the Pb geometry does not significantly degrade the signal spectrum. Per incident fusion neutron, the simulation yields 1 transmitted neutron and $11.3\times 10^{-3}$ gammas as secondaries. The gamma-to-neutron ratio at the waiting target is therefore $\sim 1\%$, meaning the signal is neutron-dominated, favorable for neutron capture measurements.

\begin{table}[h]
\centering
\begin{tabular}{l l l c c}
\hline
Simulation & Input particle & Spectrum & Neutrons & Photons \\
\hline
Background~ & Deuteron & Experimental \cite{Jia23} & 
4.5 & 1.8 \\
%per-deuteron numbers: $1.55\times10^{-3}$ & $0.60\times10^{-3}$ \\
Background & Deuteron & Maxwellian (PIC) & 
0.57 & 0.14 \\
%per-deuteron numbers: $0.19\times10^{-3}$ & $4.76\times10^{-5}$ \\
Signal     & Fusion neutron & Gaussian, $\langle E_n\rangle=2.45$ MeV & $1$ & $0.01$ \\
\hline
\end{tabular}
\caption{\label{tab:signal_background} Relative contributions of signal and background neutrons and photons at the neutron waiting target, normalized to fusion neutron number.  }
\end{table}

%Spherical Geant4 geometry approximation
To compare the total particle yields across signal fusion neutrons, secondary neutrons from deuteron scattering and all sources of photons, we normalize the GEANT4 yields to the number of incident neutrons in \tab{signal_background}.  In the conservative approximation that the deuteron distribution is also isotropic, we have from the simulations $\simeq 3\times 10^3$ deuterons in the hard distribution are accelerated for each fusion neutron produced.
The signal neutron yield per fusion neutron exceeds the background neutron yield per deuteron for the simulation-derived deuteron distribution but not for the higher energy distribution seen on the TPW. The simulation-derived Model 1 (Maxwellian) gives the more optimistic background estimate, while the TPW experiment-derived model 2 provides the conservative bound.  However, these preliminary simulations were conducted in the conventional spherical geometry model, which simplifies source and stopping analysis.  We expect the signal-to-noise to be improved over both models by situating the neutron-waiting material at $90^\circ$ to the laser axis and/or plasma surface normal axis, because past experimental results and three-dimensional simulation of the laser-plasma interaction show the highest energy deuterons are forward-directed in a $<45^\circ$ cone.  This estimate is designed to err conservative.

The GEANT4 transport simulation is also used to determine the optimal placement of neutron time-of-flight (TOF) detectors at the PHELIX target area. TOF spectroscopy reconstructs the neutron energy from the measured time difference $\Delta t_{\gamma-n}$ between the prompt gamma signal and the delayed neutron signal at a known flight distance L. The energy resolution improves with increasing flight path, making the choice of detector distance a key experimental design parameter.

\begin{figure}
\centering
\includegraphics[width=0.6\textwidth]{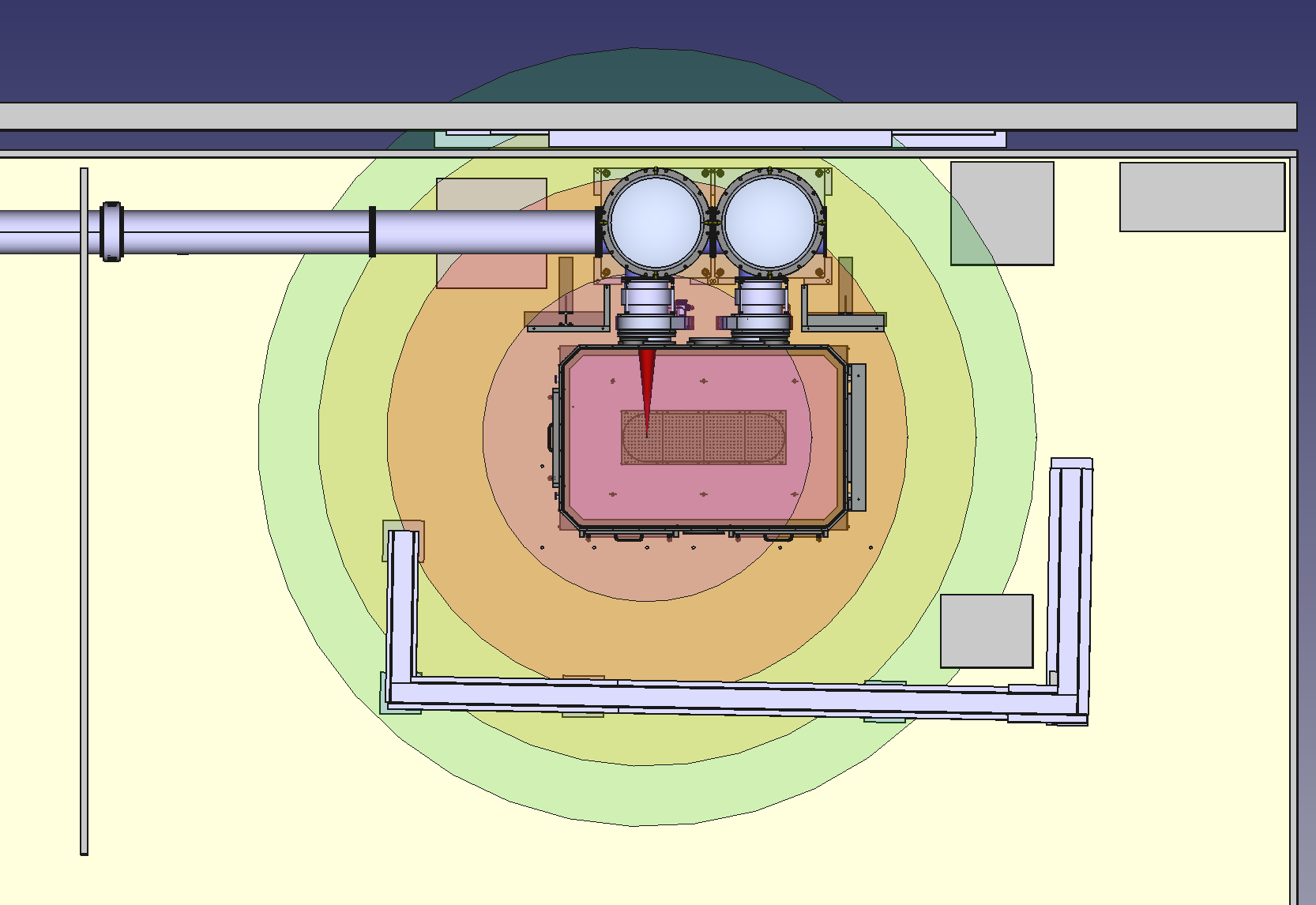}
\includegraphics[width=0.7\textwidth]{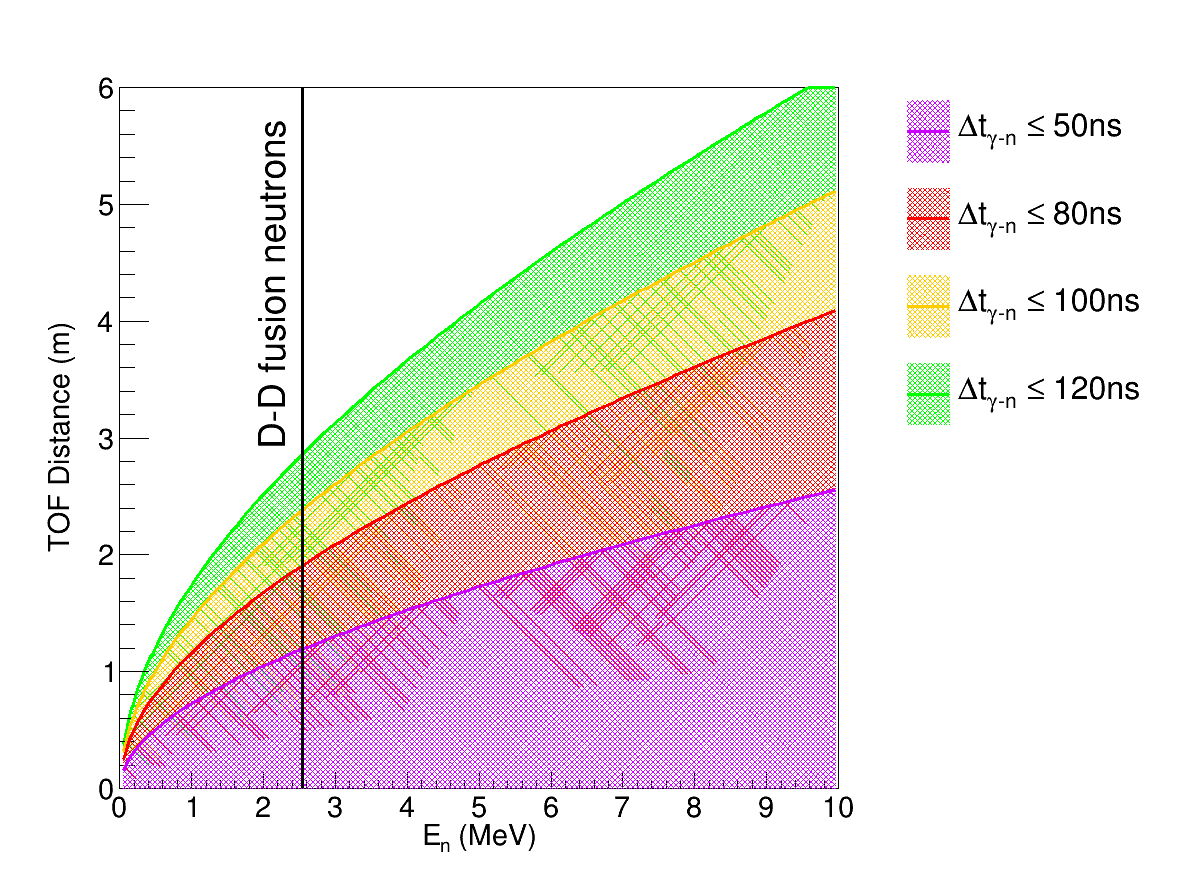}
    \caption{\label{fig:PHELIX_TOF} TOF required to achieve $\gamma$-neutron time separations of 50-120 ns across the 0-10 MeV energy range.  The minimum distance (vertical line) is derived for DD fusion neutrons at 2.45 MeV.}
\end{figure}

Figure \ref{fig:PHELIX_TOF} shows $\Delta t_{\gamma-n}$ as a function of neutron energy for four representative time differences (50, 80, 100, and 120 ns) across the range 0–10 MeV. To effectively reconstruct the DD fusion neutron spectrum up to 2.5 MeV, the simulation shows that a minimum flight path of $L \geq 2.4$ m is required between the interaction point and the TOF detectors. This corresponds to a gamma-neutron time separation of at least $\sim 80$ ns at 2.45 MeV, sufficient to separate the neutron signal from the prompt gamma flash. For neutrons at higher energies, which may be present based on the 2017 TPW experiment results, larger TOF distances are preferable to maintain adequate energy resolution. Practical implementation is subject to the spatial constraints of the target area geometry, such as visible in our model of the layout at PHELIX (\fig{PHELIX_TOF} left).

\subsubsection{Laser wakefield accelerator photoneutron signal}

\begin{figure}[h!]
\centering
\includegraphics[width=0.48\textwidth]{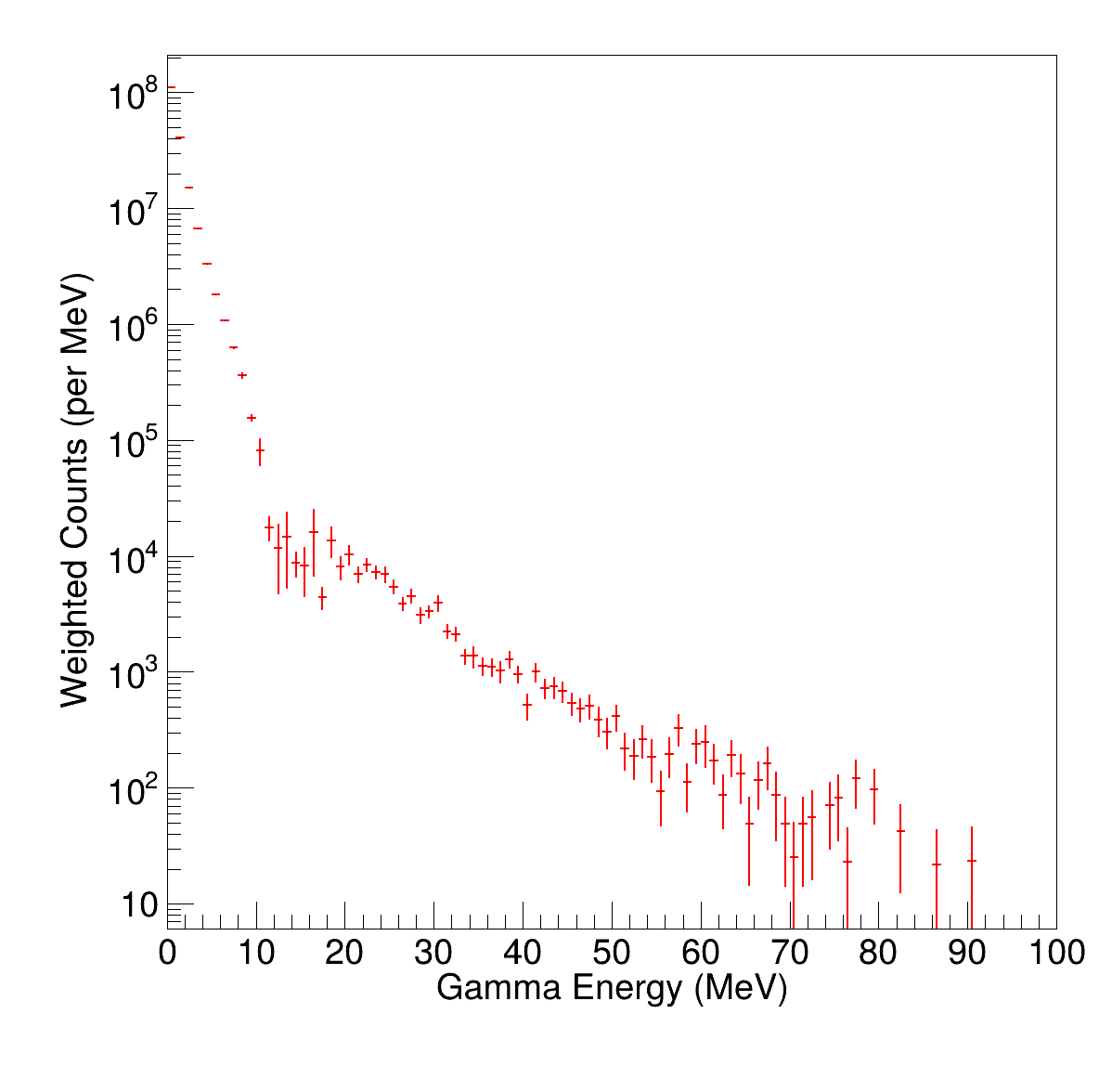}
\includegraphics[width=0.48\textwidth]{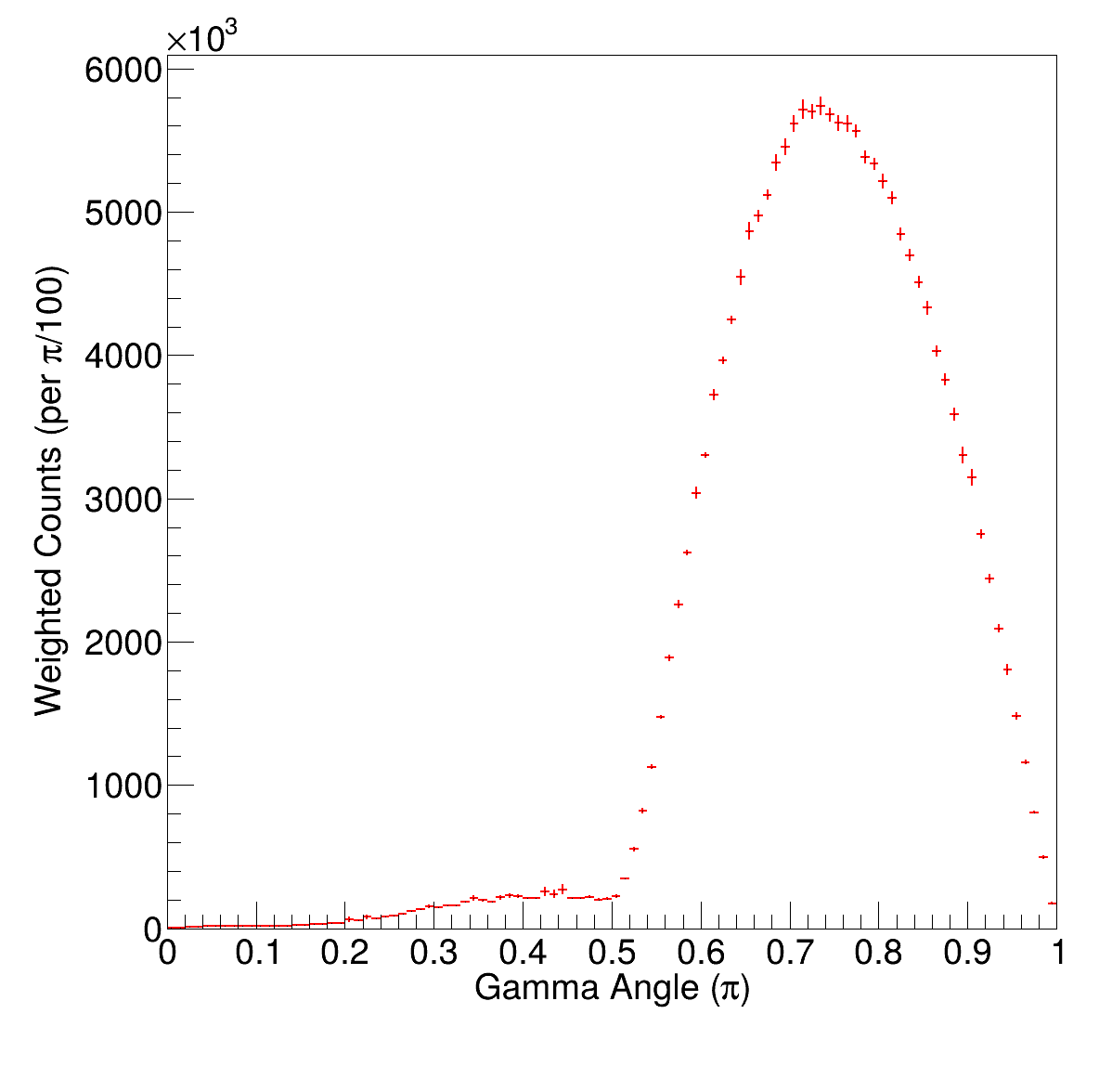}
    \caption{\label{fig:UT3_brem} GEANT4-simulated bremsstrahlung photon source characteristics from UT3 LWFA electron beam on tungsten converter.  Left: angular distribution peaking near 0.65-0.7$\pi$.  Right: energy spectrum extending to $\sim 90$ MeV with an approximately exponential shape above threshold.}
\end{figure}

For UT3 conditions, the bremsstrahlung photon energy spectrum produced by the LWFA electron beam in the tungsten converter is shown in \fig{UT3_brem}. The spectrum is approximately exponential, rising steeply at low energies $<10$ MeV and extending to the endpoint of the electron spectrum at $90\,\mathrm{MeV}$. The photon yield decreases by roughly four orders of magnitude between $1\,\mathrm{MeV}$ and $30\,\mathrm{MeV}$, with a noticeable structure near $8$--$10\,\mathrm{MeV}$ corresponding to the photonuclear giant dipole resonance threshold in tungsten.  The angular distribution of the bremsstrahlung photons, also shown in \fig{UT3_brem}, peaks near $0.65$--$0.7\pi$ and drops sharply toward 0, which is due to the rectangular prism converter geometry scattering more of the photons in the forward direction, while high angle (also lower energy) photons emitted near the edges of the converter are able to escape.

\begin{figure}[h!]
\centering
\includegraphics[width=0.48\textwidth]{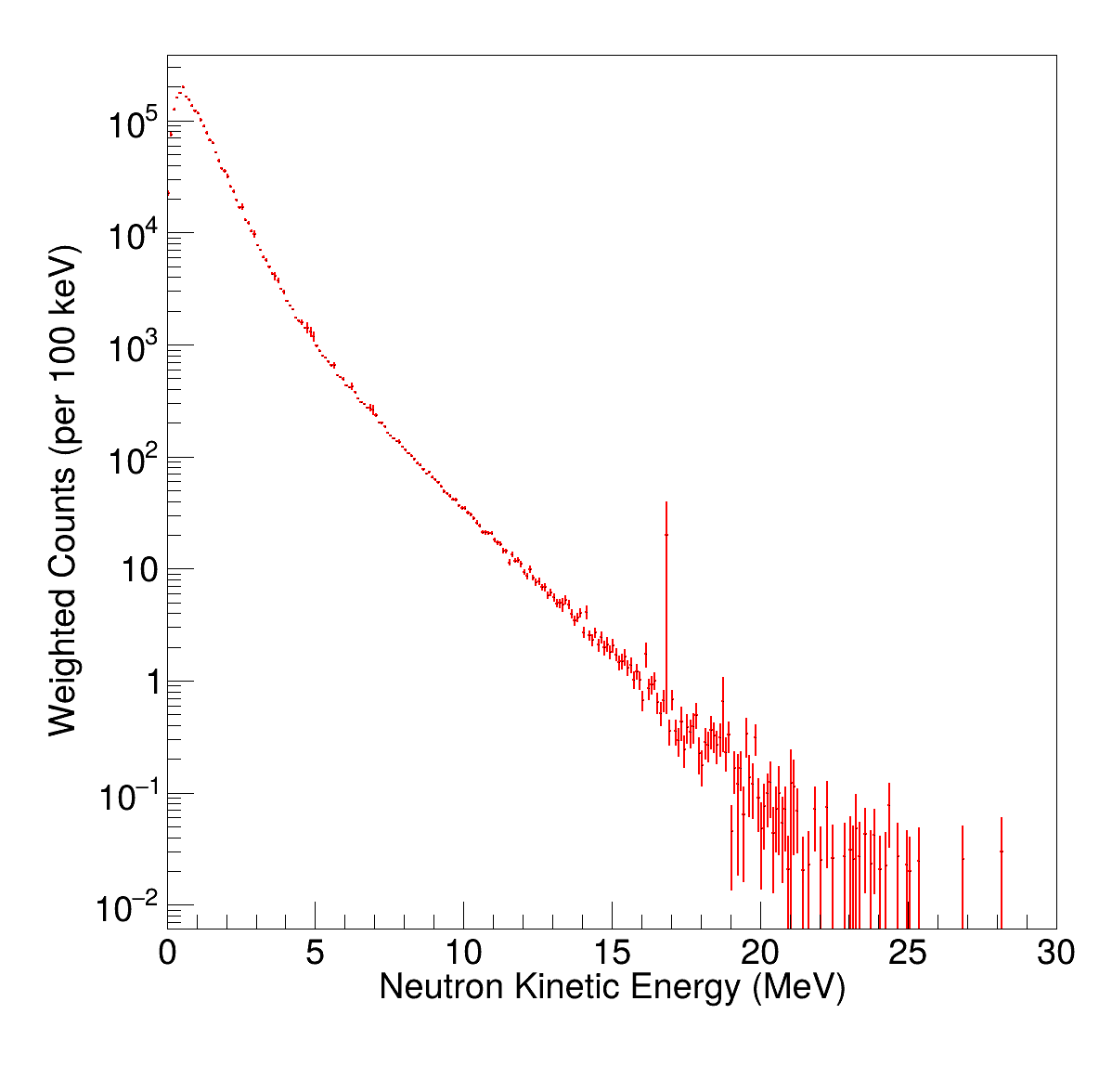}
\includegraphics[width=0.48\textwidth]{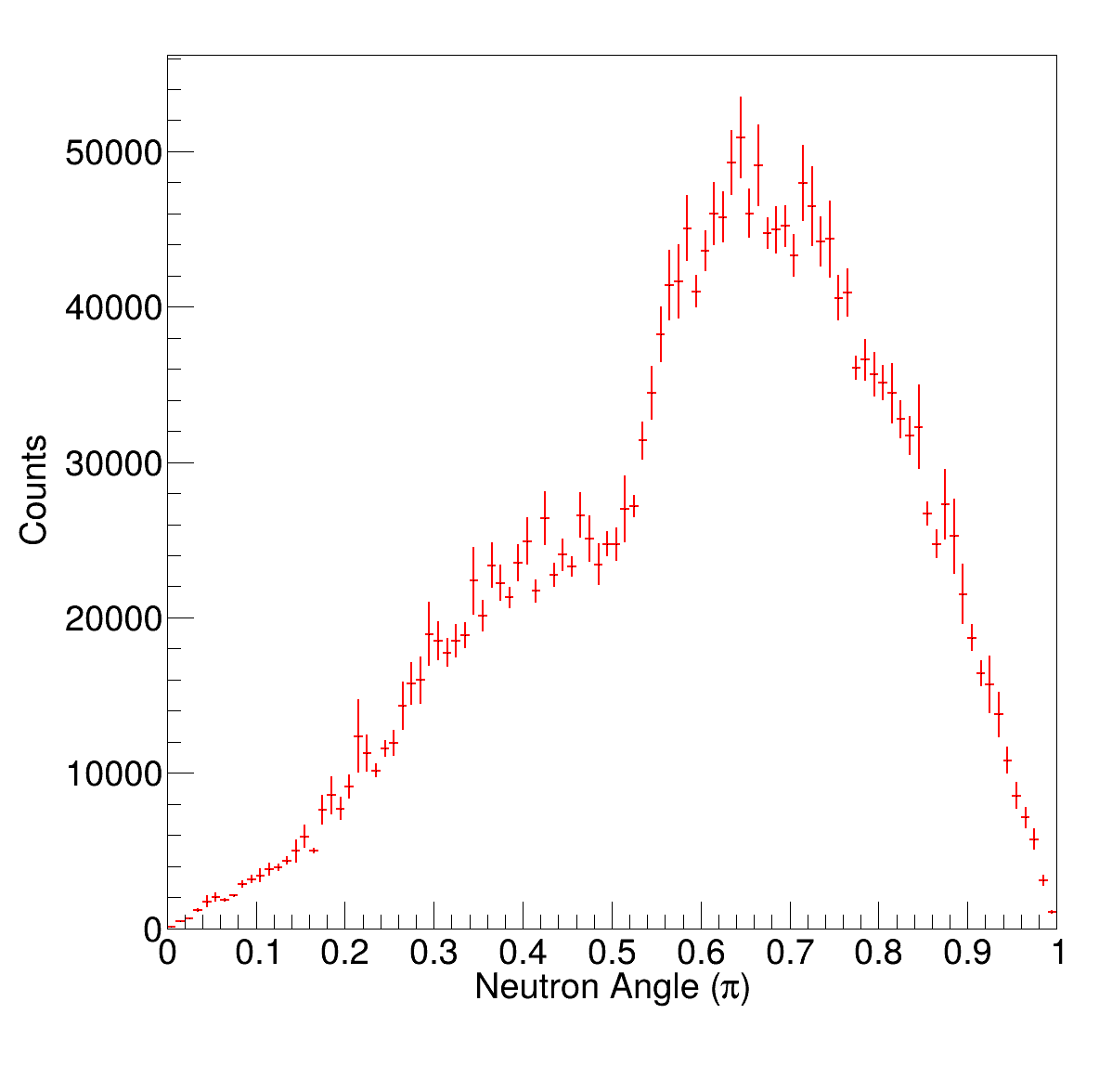}
    \caption{\label{fig:UT3_n} GEANT4-simulated photonuclear neutron source characteristics.  Left: angular distribution peaking near 0.6-0.65$\pi$ and broader than the photon distribution.  Right: kinetic energy spectrum peaking below 1 MeV extending to $\sim 28$ MeV with an approximately exponential shape above threshold.}  
\end{figure}

The resulting photonuclear neutron energy spectrum is shown in \fig{UT3_n}. Similar to the photon spectrum, it is approximately exponential, peaking strongly below $1\,\mathrm{MeV}$ and decreasing by roughly four orders of magnitude by $\sim 10\,\mathrm{MeV}$, with a high-energy tail extending to approximately $28\,\mathrm{MeV}$. The neutron angular distribution peaks near $0.6$--$0.65\pi$ and is broader than the photon angular distribution. This behavior is consistent with the near-isotropic redistribution of neutron momenta by photonuclear reaction kinematics combined with the rectangular geometry of the converter.  

\begin{figure}[h!]
\centering
\includegraphics[width=0.48\textwidth]{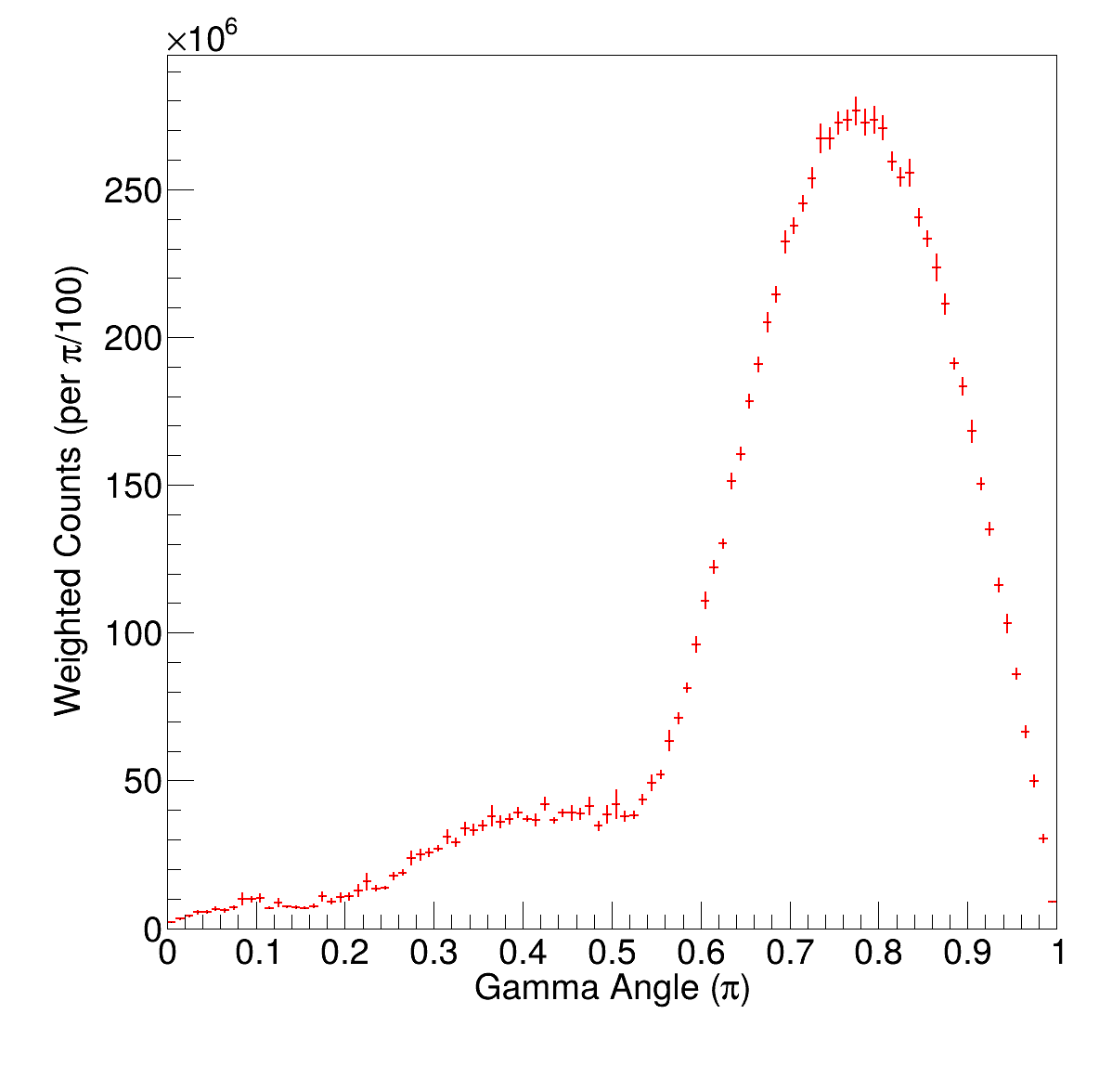}
\includegraphics[width=0.48\textwidth]{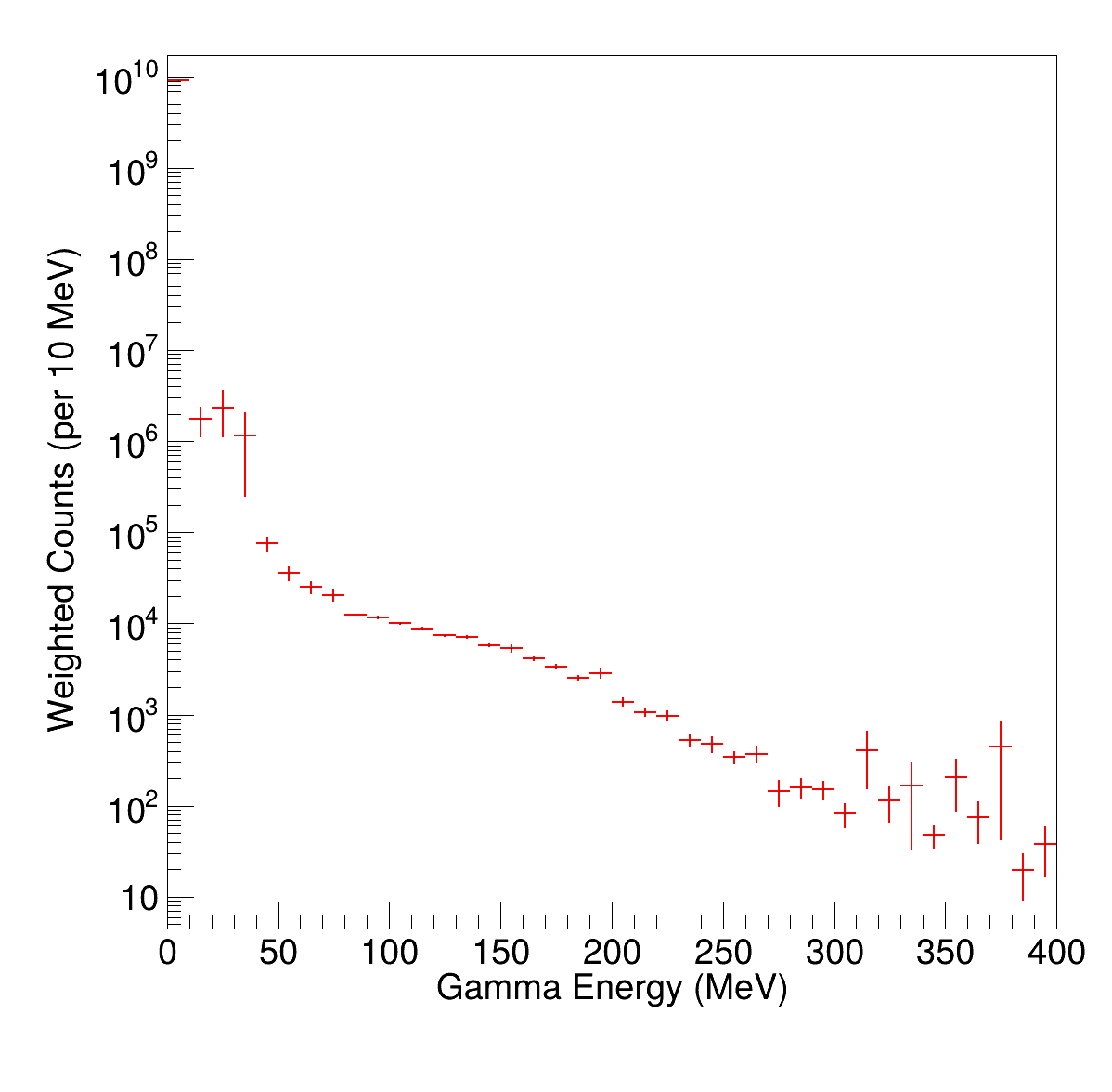}
    \caption{\label{fig:TPW_brem} GEANT4-simulated bremsstrahlung photon source characteristics from TPW LWFA electron beam on tungsten converter.  Left: angular distribution peaking near 0.65-0.7$\pi$.  Right: energy spectrum extending to $\sim 90$ MeV with an approximately exponential shape above threshold.}
\end{figure}

\begin{figure}[h!]
\centering
\includegraphics[width=0.48\textwidth]{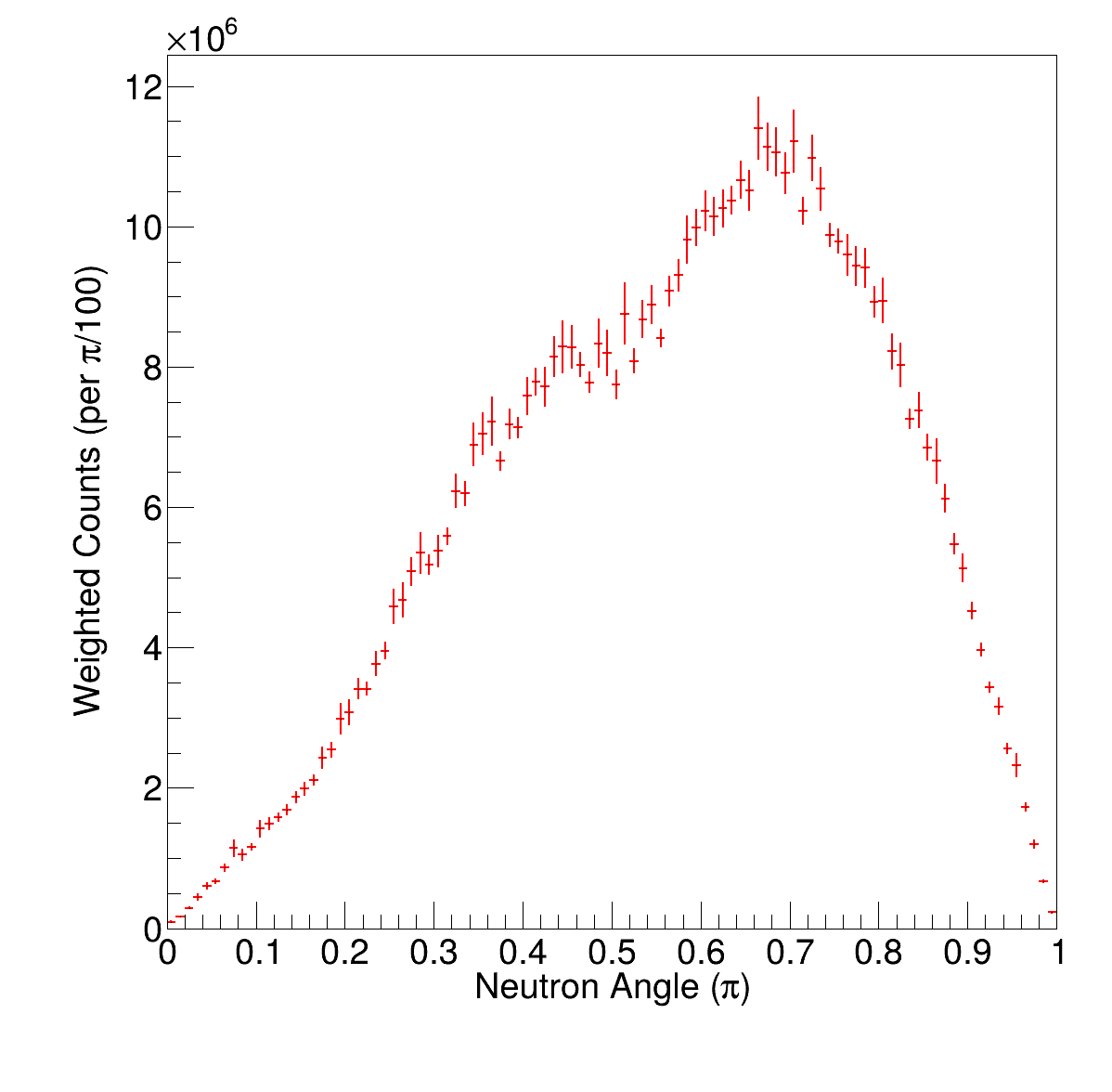}
\includegraphics[width=0.48\textwidth]{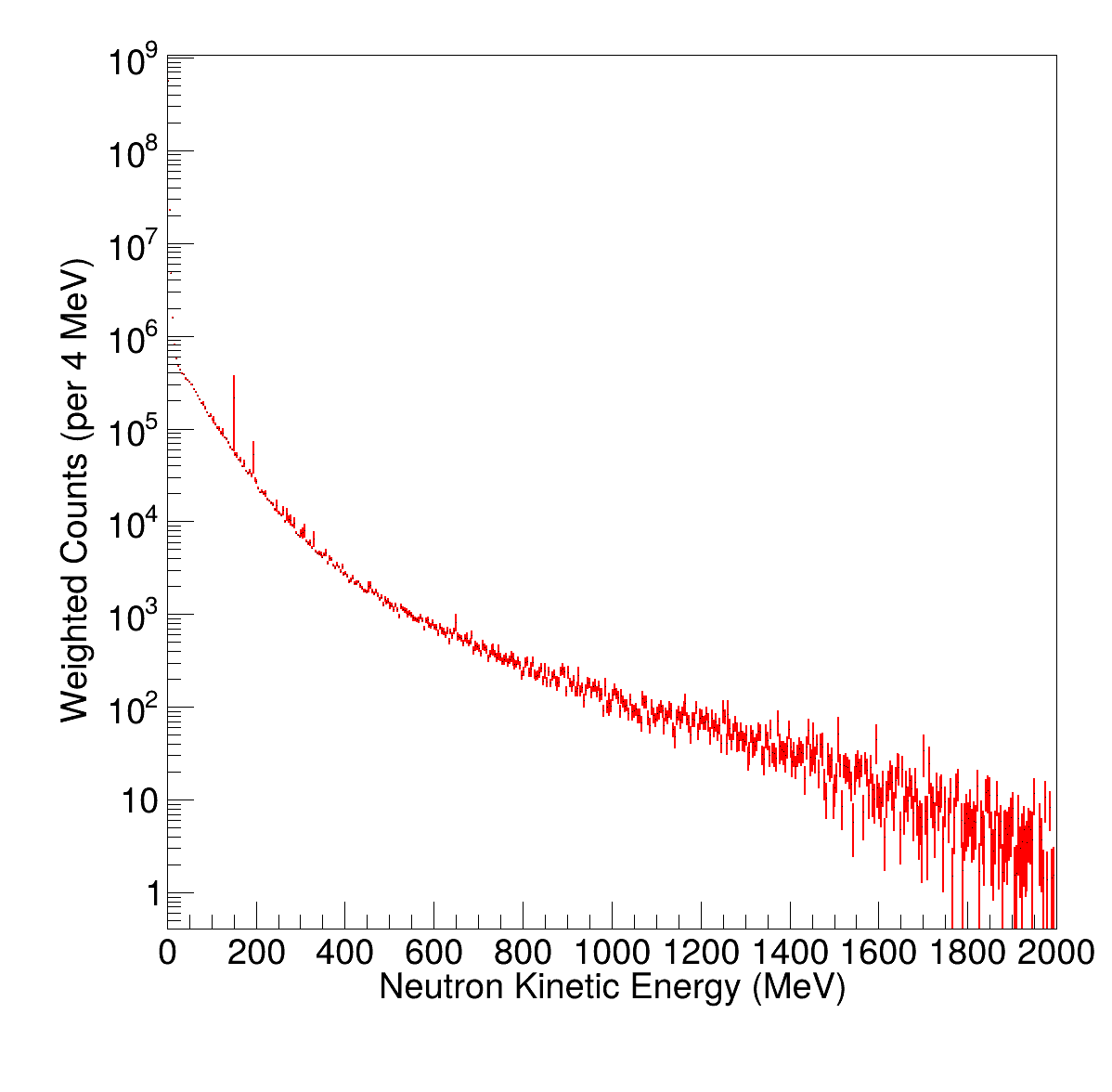}
    \caption{\label{fig:TPW_n} GEANT4-simulated photonuclear neutron source characteristics.  Left: angular distribution peaking near 0.6-0.65$\pi$ and broader than the photon distribution.  Right: kinetic energy spectrum peaking below 1 MeV extending to $\sim 28$ MeV with an approximately exponential shape.}  
\end{figure}

Repeating the simulations for the TPW's GeV-scale electron beam, we see similar shapes in the bremsstrahlung photon spectrum and angular distributions, shown in \fig{TPW_brem}.  The spectrum extends to much higher energy: the endpoint is not shown in order to resolve the peak feature in the GDR energy region.  The corresponding spectrum and angular distributions of neutrons are shown in \fig{TPW_n}.  The neutron spectrum extends to 2 GeV, with a change in the power law around 400 MeV due to the increasing importance of direct electron-nucleus processes contributing to neutron production.  These very high energy neutrons are found in the forward hemisphere, with $E_n>500$ MeV all having $\theta<\pi /2$.  The remaining neutrons are again more isotropic, with the highest number observed transverse to the beam direction due to the shorter path length out of the converter.

\subsection{Stage 3: NON-SMOKER Monte Carlo for multi-neutron capture}
\label{sec:stage3_nonsmoker}

The third stage converts the neutron fluence at the waiting targets into rapid neutron-capture event yields using an in-house Monte Carlo event generator built on the \textsc{NON-SMOKER} statistical-model cross-section database and evaluated nuclear data from ENDF. We adapt the framework of Ref.~\cite{Hil21} to our experimental geometry, target conditions, and seed nuclides. Given the incident neutron spectrum and fluence from Stage~2, the code calculates the probabilities of single- and multi-neutron capture events on $^{197}\mathrm{Au}$ and $^{103}\mathrm{Rh}$ on a per-laser-shot basis.

The choice of seed nuclides is motivated by specific nuclear-physics considerations. Gold ($^{197}\mathrm{Au}$) serves as an \emph{in situ} calibration standard: its well-measured neutron-capture cross section enables the delivered neutron fluence to be determined independently of simulation, providing an experimental check on all three simulation stages. Rhodium ($^{103}\mathrm{Rh}$) is selected as the primary rapid-capture candidate because $^{104}\mathrm{Rh}$ ($t_{1/2}=42\,\mathrm{s}$) possesses a metastable excited state, $^{104\mathrm{m}}\mathrm{Rh}$ ($t_{1/2}=4.3\,\mathrm{min}$), with a substantially longer $\beta$-decay lifetime. The neutron-capture cross section on $^{104\mathrm{m}}\mathrm{Rh}$ is approximately $20\times$ larger than that on $^{104}\mathrm{Rh}$~\cite{Mug06}, making the excited state a favorable gateway for two-neutron capture.

\begin{figure}
\centering
\includegraphics[width=0.48\textwidth]{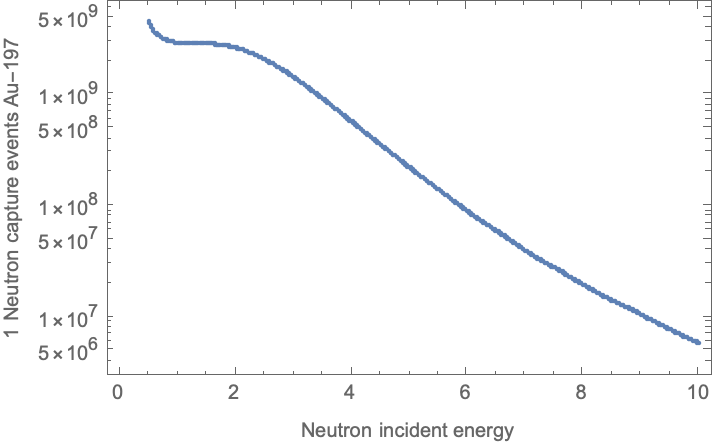}
\includegraphics[width=0.48\textwidth]{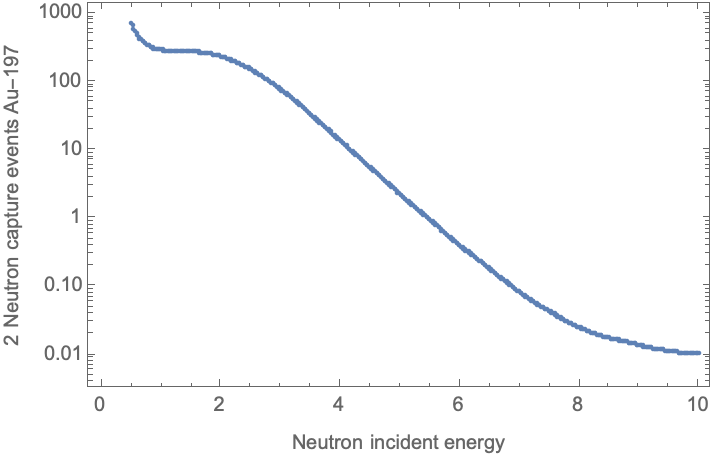}\\
\includegraphics[width=0.48\textwidth]{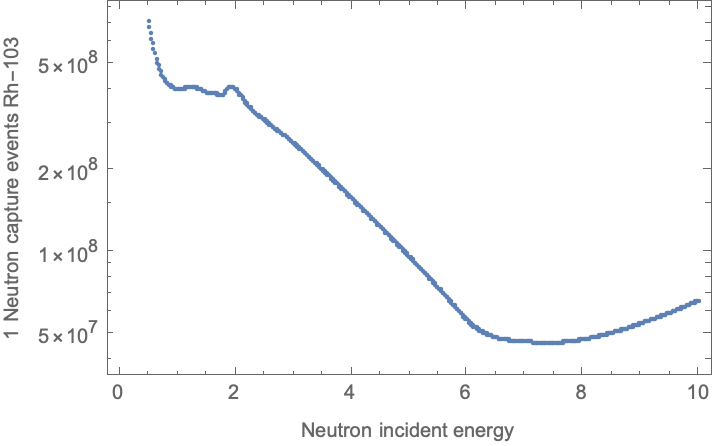}
\includegraphics[width=0.48\textwidth]{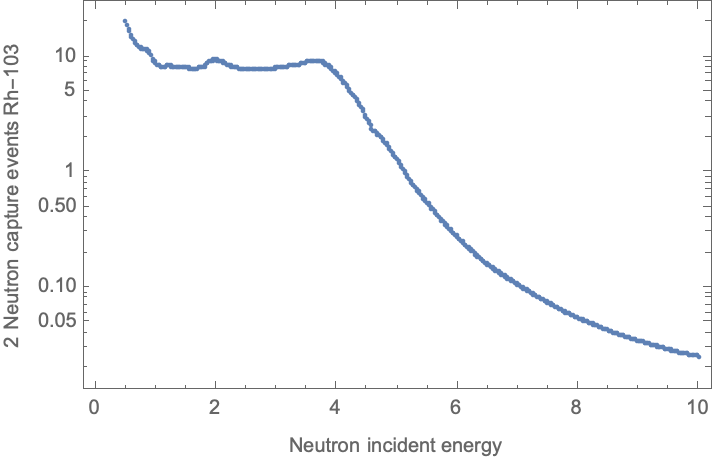}
    \caption{\label{fig:NONSMOKER1} NONSMOKER predicted 1- and 2-neutron capture yields on $^{197}$Au (upper) and $^{108}$Rh (lower) as a function of neutron energy for $10^{10}$ incident neutrons.  Recall the neutron source is expected to have an energy peaked around 1 or 2.5 MeV}
\end{figure}

The \textsc{NON-SMOKER} predictions for one- and two-neutron capture yields on $^{197}\mathrm{Au}$ and $^{103}\mathrm{Rh}$ as a function of incident neutron energy are shown in \fig{NONSMOKER1}. For the DD fusion neutron source, the peak neutron flux centered near $2.45\,\mathrm{MeV}$ falls well within a favorable region of the capture cross sections for both nuclides. Because multi-neutron capture events are intrinsically low-probability, the start-to-end simulation framework is essential for predicting the expected signal above background.

The laser repetition rate enters the calculation explicitly at this stage. Cumulative yields are computed as the product of the per-shot capture yield and the number of shots accumulated within the relevant isomer lifetime. For $^{105}\mathrm{Rh}$ ($t_{1/2}\simeq 35\,$hr),
%$^{104\mathrm{m}}\mathrm{Rh}$ ($t_{1/2}=4.3\,\mathrm{min}$), 
DD fusion at PHELIX ($\sim$1~shot/hour) contributes approximately four shots per half-life, whereas LWFA operation at $100\,\mathrm{Hz}$ contributes approximately $2.6\times10^{4}$ shots per half-life, corresponding to an advantage of roughly a factor of $6.5\times10^{3}$ in cumulative yield for short-lived isomers.

\section{Overall comparison and capture implications}

In this section, we compare the utility and physics constraints on the different neutron production methods.

\subsection{Source properties}

All three neutron source mechanisms compared in \tab{neutron_source_comparison} can achieve high conversion efficiencies of up to $10^{8}$~neutron per joule of laser energy, but differ in spectral properties and beam properties.  While the ion pitcher-catcher sources achieve high efficiency by virtue of optimizing laser-to-ion conversion efficiency, the exponential spectrum and low speed of the deuterons combine with the neutron production kinematics to produce a broad, exponential neutron energy spectrum and nanosecond pulse duration.  Also, in general the MeV ion source means multiple reaction channels are open, contributing other products to the beam and complicating spectral analysis of the outgoing neutron beam. These features make the pitcher-catcher method poorly suited to the applications requiring precision spectrum information, such as neutron capture measurements or time-of-flight spectroscopy.  Though the ion pitcher-catcher achieved this high level of efficiency earliest \cite{Jun13,Rot13}, bulk fusion and LWFA photoneutron methods have since reported cases achieving similar overall efficiency.  The latter have advantages in spectral and/or beam properties as well as target engineering.

\begin{table}
\begin{tabular}{c c c c c c}
Source type & Yield [n/J] & Spectrum & Pulse duration & Angular distribution & Source scale \\ 
\hline
Ion pitcher-catcher & up to $10^{8}$ & Exponential & $\gtrsim 1$ ns & Isotropic, but common & 1 cm \\[-3mm]
 & & & & forward component & \\
DD fusion & up to $10^{8}$ & Monoenergetic & $\lesssim 20$ ps & Isotropic & 100 \si{\micro\meter} \\
LWFA photoneutron & up to $10^{8}$ & Exponential & $\gtrsim 50$ ps & Isotropic (MeV), & $\gtrsim 1$ cm \\[-3mm] & & & &  beamed ($\gtrsim 400$ MeV) &
\end{tabular}
\caption{\label{tab:neutron_source_comparison} Summary of neutron source properties}
\end{table}

LWFA photoneutron sources also generate an exponential neutron spectrum, but deliver ultrashort pulse durations thanks to the relativistic speed of the electrons and photons transiting the converter ($\tau_n \simeq L/c$).  The pulse duration could be decreased at the cost of decreasing yield, but have an effective lower bound once the length drops much below the radiation length of the converter material (eg 3.5 cm for tungsten), as exhibited for example by Ref \cite{Pom14}.  The optimal compromise between yield and pulse duration is expected to be converter length between 1 and 2 radiation lengths.  High conversion efficiencies become possible because the LWFA beam does not need to be high quality: the accelerator can be optimized for high charge with centroid energy of order 100 MeV, which suffices for high photon yield in the GDR region.  The recent report of $5\times 10^9$ n/sr \cite{Gun22} and short pulse duration make LWFA-driven sources competitive in peak flux.  On the other hand, many photons that do not convert to neutrons can pass out of the converter contributing background in the neutron-waiting material.  Utilizing gas jets to convert laser energy into particle energy, LWFA repetition rates are currently limited by the drive laser repetition rate, gradually improving with recent reports at the 100 Hz \cite{Tau25} and 1 kHz \cite{Laz24} level.  These features of short pulse duration, high flux, and high repetition rate make the LWFA a good choice for systematic studies requiring fast neutron irradiation.  

DD bulk fusion has the most advantages in beam and source properties other than yield.  It produces the shortest pulses ($\lesssim \unit{20}{ps}$) and a distinct, quasi-monoenergetic spectrum centered near 2.45 MeV with a narrow thermal width, which are the best available for time-of-flight (TOF) spectroscopy or rapid capture with fast neutrons.  However DD fusion is the least well-characterized and optimized experimentally, because increasing the neutron yield is the most difficult: the laser should have both high intensity $a_0\gg 1$ and high energy, currently only seen in low repetition rate glass lasers, such as TPW and PHELIX.  The laser energy and neutron yield could be increased at fixed intensity by increasing pulse duration up to the picosecond scale, at which point plasma expansion dynamics begin to dominate and cause the neutron production rate to decrease.  In this regard, PHELIX at GSI and the Titan laser at JLF (\tab{laser_facilities}) should be able to match or exceed the TPW neutron yield and flux but call for further study on focusing and target engineering to optimize laser energy coupling into the plasma.   A higher repetition rate laser in this class would push required investment in target engineering to provide deuterated material at the required repetition rate.

\subsection{Transport considerations}
Geometrical attributes, including the physical source scale and angular distribution, determine the transport and divergence and hence the advantages or limitations in delivering the neutrons to an object or material to be probed. Both pitcher-catcher and LWFA platforms require spatial footprints of $\sim 1$ cm or greater due to conversion processes from ion or electron kinetic energy into neutrons. Pitcher-catcher systems at least have the advantage of partial collimation, having pronounced forward components with neutron energies in the few to 10s of MeV and divergences of $20^\circ$ to $40^\circ$.  The combination of good yield and high divergence make this approach advantageous for neutron radiography.  The best layout with this source is to place the interrogation target forward of the converter, but the pulse duration remains geometry-limited upwards of nanoseconds due to the need to separate the catcher from the pitcher and stop all ions and other particles in the catcher.  

LWFA photoneutron systems have centimeter-scale dimensions driven by electron stopping ranges and photon scattering inside a thick converter target.  Nevertheless with high charge ($>$ nC) electron beams now available from LWFAs, high peak fluxes are possible at the surfaces of the converter.  Taking advantage of the converter geometry and almost isotropic flux in the MeV energy range, neutron receiving targets can be placed effectively at the transverse surface of the converter to maximize the flux.  However additional Monte Carlo simulations should study what shielding would be required to suppress the large MeV photon background coming with the neutrons.  Multi-GeV LWFAs offer access to roughly collimated beams of very high energy ($\gtrsim 400$ MeV) neutrons due to the relativistic $1/\gamma$ beaming inherited from the core electron beam and shower development.  Like the pitcher-catcher case, this beaming considerably simplifies transport and utilization with a neutron-interrogation target placed directly at the far end of the converter to maximize yield.

DD bulk fusion stands out as an exceptionally compact source with a spatial scale of $\sim 100 \si{\micro\meter}$, while providing an inherently isotropic angular distribution that simplifies multi-angle TOF diagnostics.  The fusion source offers the most narrowly defined neutron source simplifying transport and shielding design.  High energy deuterons generated by the hole-boring dynamic are strongly forward-directed, so we should be able to largely preclude the most dangerous $>10$ MeV deuterons from contributing to background simply by situating the neutron-interrogation target at $\sim 90^\circ$ to the laser direction.  Since such high energy deuterons are in fact undesirable, we expect that further effort optimizing the combination of laser intensity and target density profile can find an optimum where hole boring is slow.  Recall that a majority of the energy is transferred to quasi-thermal sub-MeV deuterons via dissipation of the laser-driven channel, so the target of optimization is to extend this hole boring for as long as possible thereby increasing total energy deposited.

\subsection{Neutron capture yields}

Maximizing the 2-neutron capture yields depends two additional facility-level and experiment-level considerations.  First, the total 2-neutron event yield depends on the product of two parameters of the neutron source: the number of neutrons delivered to the target material per shot $N_n$ and the number of shots possible per half-life of the desired product state, which we write as the product of the accelerator repetition rate and isotope half-life, $\nu_{\rm acc}t_{1/2}$.  We maximize the signal for post-exposure isotope identification by integrating neutron exposure over an interval $1.5t_{\rm 1/2}$.  Writing $P[A\to A+2]$ for the per-neutron probability of 2-neutron capture, the product $1.5N_n\nu_{\rm acc}t_{1/2}P[A\to A+2]$ estimates the number of events.  Clearly higher repetition rate offers access to a wider range of neutron rich nuclei.

We can now directly compare the effectiveness of a 1-J LWFA neutron source to 100-J DD fusion source in delivering capture events.  The 1-J LWFA photonuclear source provides $\sim 10^7 - 10^8$ neutrons per shot and can operate at 1 Hz.  The 100-J DD fusion source can deliver $\sim 10^9-10^{10}$ neutrons per shot, but operates at $1/3600$ Hz.  Hence the ratio of captures events
\begin{align}\label{eq:LWFAvsDD2ncapture}
    \frac{Y^{(LWFA)}}{Y^{(DD)}}\simeq \frac{N_n^{(LWFA)}\nu_{\rm acc}^{(1J)}}{N_n^{(DD)}\nu_{\rm acc}^{(100J)}}\simeq 36
\end{align}
favors the LWFA in the near term.  From \fig{NONSMOKER1}, we observe that differences in the delivered neutron spectrum should make only a small difference in the yield: the exponential spectrum of LWFA has $\gtrsim 70\%$ of its weight in the 1-2 MeV where the 2-neutron yield curves are roughly flat.  Advancements in repetition rate are expected in both classes of laser, with 100-Hz, 1-J systems under construction and 0.1-1 Hz PW-class systems nearly ready for users.  If both of these laser technologies are realized, the ratio \eq{LWFAvsDD2ncapture} will become closer to 1, with the $\sim 100\times$ higher yield achieved by DD fusion compensated by $\sim 100\times$ higher repetition rate of LWFA sources.  

Clearly, these two facility types retain some complementarity.  DD fusion offers higher on per-shot flux, controlled spectrum and background.  The high per-shot flux may give access to very short-lived metastable states for multiple capture but the lower repetition rate reduces the number of shots that can integrated toward the signal in the multi-capture end-state isotope.  LWFAs offer rep-rate for integration, good for moderately short-lived isomers and accessibility at tabletop facilities. Neither is universally superior.  The optimal choice depends on the target isomer's half-life and the available facility.  These comparisons are summarized in \tab{neutron_sources}.

\begin{table}[h]
\centering
    \begin{tabular}{lcc}
    \hline
    Property & DD Fusion (PW) & LWFA Photonuclear (TW) \\
    \hline
    Laser energy               & $\sim$100--250 J      & $\sim$1 J \\
    Laser regime               & PW   & 10s TW \\
    Neutron yield/shot         & $10^9-10^{10}$           & $10^7-10^{8}$ \\
    %Neutrons/joule             & $\sim10^{7}$         & [number] \\
    Neutron energy             & $\sim$2.45 MeV (narrow) & Broad, harder spectrum \\
    Angular distribution       & Isotropic      & Isotropic \\
    Dominant background       & Neutron     & Photon \\
    Rep rate (future)       & 1/hr (1 Hz)        & 1 Hz  (100 Hz)\\
    2n-capture/shot ($^{103}$Rh) & $0.8-8$       & $0.008-0.08$ \\
    2n-capture/hour ($^{103}$Rh) & 8            & 290 \\
    Best facility              & PHELIX, ELI-NP       & UT3, Tau Labs \\
   % Scaling law                & $\propto E_{\mathrm{laser}}^{\alpha}$ (fusion) 
    %                            & $\propto E_{\mathrm{laser}}^{\beta}$ (LWFA, different $\beta$) \\
    \hline
    \end{tabular}
    \caption{\label{tab:neutron_sources} Comparison of DD fusion and LWFA photonuclear neutron characteristics from laser to neutron source to capture event estimate.}
\end{table}

\begin{figure}
\centering
\includegraphics[width=1.0\textwidth]{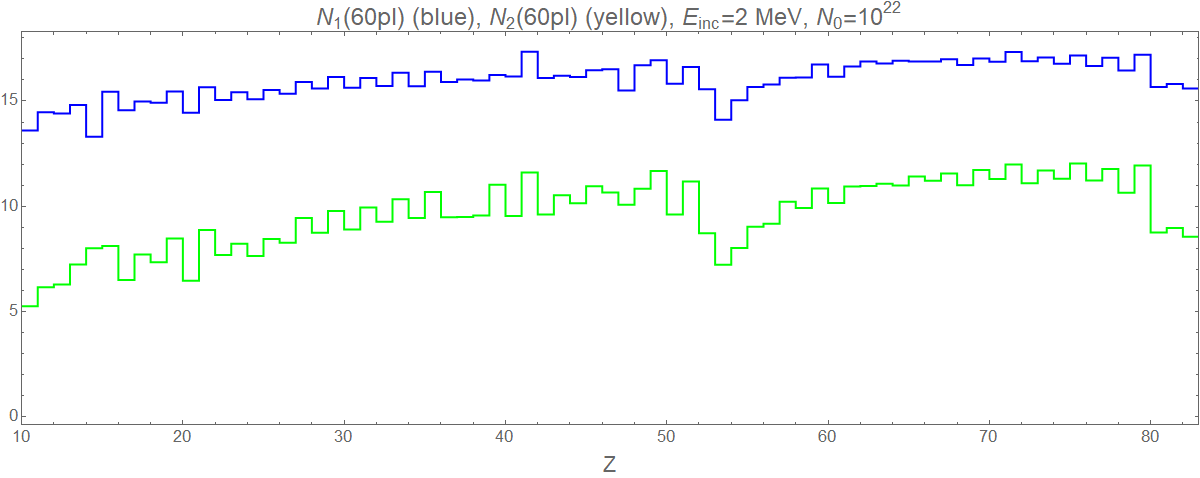}
    \caption{\label{fig:NONSMOKER2} Prediction of 1- (blue) and 2- (green) neutron capture yields as a function of atomic number $Z$, using the heaviest isotope with $t_{1/2}>1$ hr.  Neutron flux $10^{21}$/cm$^2$/s, effective neutron number (neutrons/shot times number of shots) $N_n^{(tot)}=6\times 10^{18}$, neutron energy 2 MeV.  Vertical axis is log scale.}
\end{figure}

We considered 2-neutron capture on $^{103}$Rh, because the $A+2$ state is longer-lived than the $A+1$ state.  The half-life of $^{105}$Rh is 35 hours allowing for integration over many shots even on low-repetition rate ($\sim 1$/hour) PW-class lasers such as PHELIX or TPW.  Figure \ref{fig:NONSMOKER2} shows the predicted one- and two-neutron capture yields across atomic number $Z$ at a neutron flux of $10^{21}\,\mathrm{cm^{-2}\,s^{-1}}$ with $10^{9}$ incident neutrons at $2\,\mathrm{MeV}$, accumulated over 60 shots at a repetition rate of $1\,\mathrm{Hz}$. These results demonstrate that detectable yields are achievable across a broad range of seed nuclides. This comparison directly informs the facility-selection discussion in Sec.~IV.

%Sensitivity to simulation assumptions:
%Deuteron spectrum shape (exponential vs. Maxwellian): $8\times$ difference in background
%2D vs. 3D PIC normalization uncertainty

%NON-SMOKER cross section uncertainty (factor 2–10 known)
%Background discrimination: The qualitatively different background character of the two sources (neutron-dominated for DD, gamma-dominated for LWFA) suggests different experimental strategies for background rejection: discuss.
The simulation scheme presented here is by no means exhaustive.  We have identified essential aspects of the physics where dedicated investigation is desirable to optimize overall performance and ensure accurate data is passed to the next stage.  Some physics will only be validated by experiment: PIC simulations in the high density regime offer a lot of insight but rarely reproduce in detail observed accelerated particle distributions.  In our case, we are sensitive to the uncertain background from high energy deuterons in the DD fusion case.  The photon background coming with the LWFA photoneutrons can be determined with greater confidence, but we have not yet studied mitigation or its effect on the signal.  The neutron capture yield Monte Carlo presented in \sec{stage3_nonsmoker} has factor 2-10 uncertainties from the underlying cross sections, which are anyway the goal of measurement.  

%Path to experiment: TOF detector placement (≥2.4m for DD fusion spectrum reconstruction); Au-197 in-situ calibration for both sources; diagnostics differences between the two platforms.

\section{Conclusion}

%Two mechanistically distinct laser-driven neutron source types compared via complete start-to-end simulation chains
We have briefly reviewed the physics of various laser-driven neutron sources especially to understand their potential and limitations as future tools of neutron science.  We have evaluated in detail the relevant physics that needs to be accurately modeled and extracted in complete start-to-end simulations.  This allows us to identify where further improvements are possible and where physics limitations are (or close to) saturated.  This work establishes the first-round simulation framework and design criteria for the first direct laser-driven rapid neutron capture measurement, as well as prospective campaigns at PHELIX or ELI-NP for DD fusion and at UT3 or Tau Labs 100 Hz system for LWFA.
%The start-to-end framework developed here directly transfers to PHELIX and ELI-NP for DD fusion, and to UT3 or Tau Labs for LWFA. The examples offered here provide helpful guides and preliminary estimates to start the specialized simulations that a dedicated experimental campaign would entail.

%\LLB{Maybe still add some: Key quantitative comparison results: [your findings on yield, scaling, capture rates]}

Pitcher–catcher sources have previously excelled in high total neutron yield, but suffer limitations in temporal and spatial resolution for the future neutron science applications.  The long pulse duration makes pitcher-catcher sources basically unsuitable for fast neutron resonance spectroscopy, resonance transmission analysis, shock thermometry, active interrogation.  Their source size means they are unlikely to achieve the required peak fluxes for novel capture cross section measurements.  Due to these limitations, we have studied the other two methods here more extensively, as they also have been less explored experimentally.

In general, DD bulk fusion sources offer the strongest advantages in applications requiring excellent temporal resolution, point-like geometry, and high instantaneous flux.  DD fusion is the most favorable option for fast neutron resonance spectroscopy (FNRS) and shock thermometry, because the ultra-short ($\tau_n \lesssim 20$ ps) pulse duration greatly improves time-of-flight energy resolution ($\propto 1/\tau_n$) and greatly reduces temporal averaging in the nanosecond-scale shock dynamics. The extremely high peak flux should allow precise probing of transient thermodynamic conditions in shocks as well.  Quasi-monoenergetic spectrum and isotropic angular distribution enable precise, multi-angle measurements, and the effectively point-like emission ($\lesssim 100$\,\si{\micro\meter} source size) minimizes spatial blurring and enhances transmission resolution across multiple angles in neutron resonance transmission analysis.  While the ultra-high intantaneous flux would be beneficial for neutron capture cross section measurements, their relatively low total yield can limit statistical reach until better optimized.

LWFA photoneutron sources are the most versatile option, offering advantages in broad-spectrum interrogation, directional beams, and high repetition rate operation.  LWFA sources offer an alternative to DD fusion for FNRS by delivering ultrashort pulses combined with a broad, exponential neutron energy spectrum extending to tens of MeV. While the longer pulse slightly reduces ToF resolution compared to DD, the wide spectrum enables simultaneous probing of multiple resonance features across a large energy range in a single shot. Similarly, LWFA sources are slightly lower resolution than DD fusion but passably matched to the required timescale for dynamic compression experiments. Their advantage is a broad, continuous spectrum that allows simultaneous observation of multiple Doppler-broadened resonance features during transient events, offering richer diagnostic capability than quasi-monoenergetic sources. For NRTA, bulk radiography and active interrogation, LWFA sources sacrifice point-like spatial resolution due to their centimeter-scale source size, but driven by GeV-scale electron beams, they can compensate with forward beaming, becoming the most effective for deep-penetration imaging of dense or shielded objects. The forward-directed emission maximizes flux delivery to a target, while the short pulse duration provides a clean temporal window for separating prompt and delayed nuclear signatures.  LWFA sources turn out to be highly promising for nuclear astrophysics due to recent advances in peak flux and their broadband neutron spectrum, which allows simultaneous measurement of energy-dependent cross sections without the need for scanning. Combined with high-repetition-rate lasers, they could integrate statistics, making them well suited for studies of cross sections short-lived isotopes.

%Different scaling laws, the key physical insight: Because LWFA electron energy and charge scale differently with laser energy than deuteron acceleration efficiency, the two sources do not simply trade off linearly. Explain what this means for facility selection across the energy range 1J → 250J.
Looking forward to advances in laser technology, we note the scaling of the neutron yields with laser parameters.  The LWFA photoneutron yield scales almost linearly with laser energy, and the required increase in electron beam charge can be tuned via the plasma accelerator design.  Increasing the DD fusion yield requires retaining hole-boring level intensity across a larger spot size to increase the heated volume, also scaling linearly with laser energy, but starting already at a very laser energy and power levels.  From this point of view, the LWFA photoneutron yield has a clearer path to increase in the next generation of experiments and facilities.

%Main conclusion: the sources are complementary rather than competing — DD fusion at PW scale for maximum per-shot flux; LWFA at TW scale for high rep-rate access to short-lived isomers
From this discussion, we conclude these neutron source types are complementary tools, each optimized for a distinct region of the performance parameter space. Deuterium–deuterium (DD) fusion sources operating at petawatt (PW) scale should deliver the highest peak neutron flux per shot, making them uniquely suited for experiments requiring extreme peak brightness. In contrast, laser wakefield acceleration (LWFA)–driven photoneutron sources, typically operating at terawatt (TW) scale, offer lower per-shot yield but have advantages in broadband spectra and near-term availability of high repetition rate. This enables efficient accumulation of statistics and repeated interrogation, particularly advantageous for studies involving short-lived nuclear isomers.  Together, these source classes span a wide operational envelope, with DD fusion maximizing peak performance and LWFA systems providing flexible, high-throughput access to transient or rare phenomena.

\textbf{Acknowledgments.} This material is based upon work supported by the U.S. Department of Energy, National Nuclear Security Administration under Award Number DE-NA0004201, by the National Science Foundation under Grant Number NSF2108921; by the Air Force Office of Scientific Research under Award Number FA9550-25-1-0286; and by TAU Systems under Sponsored Research Agreement UTAUS-FA00001488.

The data that supports the findings of this study are available from the
corresponding author upon reasonable request.

\bibliographystyle{apsrev4-2}
\bibliography{ldns}

\begin{appendix}
\section{Texas Petawatt laser parameters and neutron production history}\label{appx:TPW}
The Texas Petawatt (TPW) laser facility at the University of Texas at Austin was a first generation, user-oriented, high-energy ultrashort-pulse driver for high-intensity laser-plasma experiments. Since commissioning in 2003, TPW supported both method development (diagnostics and target platforms) and physics campaigns spanning high-energy-density science, laboratory astrophysics, plasma-based particle acceleration, and secondary-source generation, with an emphasis on quantitative benchmarks of laser-plasma coupling and source performance.

TPW was an ultrashort, high-energy mixed glass laser system that combined optical parametric chirped-pulse amplification (OPCPA) with power amplification in neodymium-doped phosphate and silicate media \cite{Mar12}. Standard operation delivered ~140 J in 140 fs on target with $<5\%$ shot-to-shot energy stability, corresponding to petawatt-scale peak power. The laser operated at 1057 nm with vertical polarization and the compressed temporal profile was approximately Gaussian. Typical focused-beam quality was characterized by a Strehl ratio of $\sim0.7 $ with $\sim 10\%$  RMS shot-to-shot variation.

TPW was a single-shot system with the ability to shoot into either of two target chambers with complementary focusing geometries: TC1 (fast-focus) with standard f/3 or f/1.1 focusing for tight-focus laser-solid studies, and TC2 (long-focus) with standard f/40 focusing for extended interaction lengths commonly used for gas-target platforms.  Recent upgrades enabled operation at one shot every 30 minutes, marking a major milestone for the facility. This advancement doubled the daily shot allowance, enabling higher statistical confidence through improved repeatability and more efficient parameter scans. Beyond nominal compressed operation, TPW also supported alternative pulse configurations (e.g. intentionally lengthened pulses via compressor adjustment) to tune peak intensity and match plasma/diagnostic constraints, which was especially valuable for tailoring the interaction to different target platforms and physics regimes.

With these capabilities, the Texas Petawatt facility supported a substantial experimental record spanning laser-plasma physics and related areas of research enabled by ultrashort, high-intensity laser irradiation. In underdense plasma, TPW-driven laser wakefield acceleration demonstrated GeV-class electrons, including reports of electron acceleration up to 10 GeV in a 10 cm long nanoparticle-assisted laser-wakefield acceleration (LWFA) experiment \cite{Wan23, Ani24}. TPW also produced influential results on intense high-energy photon and pair production from laser-solid interactions, including multi-MeV bremsstrahlung/$\gamma$-ray generation \cite{Hen14} and dense $e^+e^-$ pair creation in high-Z targets at $\sim 10^{21}$ \si{\watt\per\centi\meter\squared}-class intensities \cite{Lia15}. 

In high-energy secondary-source work, TPW enabled compact neutron-generation platforms with ultrashort neutron pulses and high peak flux, establishing a widely cited benchmark for laser-driven neutron sources. The three main production mechanisms employed at the TPW facility included electron driven neutron converters, ion-driven neutron converters, and fusion neutrons generated from deuterated media. A landmark TPW neutron generation result was the demonstration of an ultra-short pulsed neutron source driven by relativistic electron beams generated via laser-solid interactions and converted to neutrons in a secondary metal target. This enabled extremely high peak neutron flux ($>10^{18}$ n/cm$^2$/s) on sub-nanosecond ($<50$ ps)  timescales \cite{Pom14}. Ion driven TPW campaigns using laser-accelerated protons produced via TNSA incident on a converter material, Lithium metal slabs, generated fast neutrons ($1.6\times 10^7$ n/sr) and established a practical compact neutron source platform \cite{Str13}. Cornerstone bulk fusion experiments performed at the TPW facility focused primarily on cluster Coulomb explosions \cite{Ban13, Que18}. The most recent result used a cryogenic deuterium jet to generate $\sim 10^{10}$ neutrons with a peak flux of $>10^{22}$ n/cm$^2$/s \cite{Jia23}.

\section{Laser-Driven Neutron Source Survey\label{sec:lit_survey}}

\begin{table}[h!]
\centering
\begin{tabular}{lcccc}
\hline
Reference & Energy (J) & Pulse duration & Peak yield (n/sr) & Angular distribution / peak flux \\
\hline
Norreys (1998)        & 8--20     & 1.3 ps   & $7.0\times10^{7}$  & Relatively isotropic \\
Disdier (1999)        & 7         & 300 fs   & $8.0\times10^{5}$  & Peaked along laser axis \\
Lancaster (2004)      & up to 80  & 1 ps     & $3.0\times10^{8}$  & Relatively anisotropic \\
Higginson (2010)      & 140       & 0.7 ps   & $2.0\times10^{8}$  & Directed \\
Higginson (2011)      & 360       & 9 ps     & $8.0\times10^{8}$  & Small angular dependence \\
Willingale (2011)     & up to 6   & 400 fs   & $5.0\times10^{4}$  & Directed forward \\
Maksimchuk (2013)     & up to 6   & 400 fs   & $4.0\times10^{5}$  & Directed beam (20$^\circ$ divergence) \\
Jung (2013)           & 80        & 600 fs   & $4.4\times10^{9}$  & Anisotropic + isotropic component \\
Roth (2013)           & 80        & 600 fs   & $1.0\times10^{10}$ & Forward peaked \\
Storm (2013)          & 60        & 180 fs   & $1.6\times10^{7}$  & Forward peaked \\
Zulick (2013)         & 1.1       & 40 fs    & $1.0\times10^{7}$  & Anisotropic (6.2$\times$ forward) \\
Guler (2016)          & 80        & 600 fs   & $1.5\times10^{9}$  & Forward-directed beam \\
Kar (2016)            & $\sim$200 & 750 fs   & $9.0\times10^{8}$  & Anisotropic ($\sim$70$^\circ$ FWHM) \\
Alejo (2017)          & 200       & 750 fs   & $2.0\times10^{9}$  & Forward-peaked ($\sim$70$^\circ$ cone) \\
Kleinschmidt (2018)   & 140--175  & 0.5 ps   & $1.4\times10^{10}$ & Directed (forward max)\\[-3mm] &&&& $4\times10^{14}$ n/cm$^{2}$ \\
Curtis (2021)         & 8         & 45 fs    & $9.5\times10^{5}$  & Directed/beamed ($<10^\circ$ divergence) \\
Yao (2023)            & 80        & 0.8 ps   & $6.6\times10^{7}$  & Forward-directed \\
Yogo (2023)           & $\sim$900 & 1.5 ps   & $2.4\times10^{10}$ & Close to isotropic (forward tendency) \\
Higginson (2024)      & $\sim$100 & 650 fs   & $1.3\times10^{8}$  & Modest angular dependence \\
Leli\`evre (2024)     & 3.2       & 22 fs    & $1.4\times10^{5}$  & Forward-directed \\
Osvay (2024)          & 21 mJ     & 12.3 fs  & $9.1\times10^{1}$  & Anisotropic \\
Stuhl (2025)          & 23 mJ     & 12 fs    & $2.1\times10^{4}$  & Anisotropic\\[-3mm] &&&& $8.6\times10^{11}$ n/s \\
Wang (2025)           & 6--7      & 45 fs    & $2.0\times10^{7}$  & Directed beam ($\sim$20$^\circ$ FWHM) \\
Yao (2025)            & 45        & 22 fs    & $4.7\times10^{7}$  & Anisotropic (shifted 15$^\circ$ from normal) \\
\hline
\end{tabular}
\caption{Pitcher-catcher method and reported yields}
\label{tab:pitcher_neutron_literature}
\end{table}

\begin{table}[h]
\centering
\begin{tabular}{lcccc}
\hline
Reference & Energy (J) & Pulse duration & Peak yield (n/sr) & Angular distribution / peak flux \\
\hline
Ditmire (1999)$^*$     & 120 mJ     & 35 fs   & $8.0\times10^{2}$  & Isotropic \\
Grillon (2002)     & 800 mJ     & 35 fs   & $6.4\times10^{2}$  & Small anisotropy \\
Madison (2004)     & up to 10   & 100 fs  & $4.8\times10^{4}$  & Isotropic \\
Lu (2009)          & 5.4        & 50 fs   & $2.0\times10^{4}$  & Isotropic \\
Bang (2013)$^*$        & 120        & 170 fs  & $1.3\times10^{6}$  & Isotropic \\
Curtis (2018)      & 1.65       & 60 fs   & $2.9\times10^{5}$  & Peaked toward target normal \\
Hah (2018)         & 18 mJ      & 45 fs   & $3.5\times10^{1}$  & Isotropic \\
Quevedo (2018)     & 100        & 150 fs  & $1.2\times10^{6}$ (total)\footnote{Quevedo (2018) reports total neutron count, not n/sr; not directly comparable to other yield entries.} & --- \\
Jiao (2023)        & 90--140    & 140 fs  & $1.7\times10^{9}$  & Predominantly isotropic\\[-3mm] &&&& $>10^{22}$ n/cm$^{2}$ \\
Knight (2024)      & 8 mJ       & 40 fs   & $8.0\times10^{0}$  & Isotropic \\
\hline
\end{tabular}
\caption{Bulk fusion method and reported yields.  References with $^*$ use cluster fusion rather than solid density plasma.}
\label{tab:fusion_neutron_literature}
\end{table}

\begin{table}[h]
\centering

\begin{tabular}{lcccc}
\hline
Reference & Energy (J) & Pulse duration & Peak yield (n/sr) & Angular distribution\\[-3mm] &&&& Peak flux \\
\hline
Kitagawa (2011)$^{\ddagger}$ & 0.6       & 150 fs  & $1.6\times10^{4}$ & Isotropic \\
Pomerantz (2014)$^*$            & 90        & 150 fs  & $1.8\times10^{8}$ & Isotropic\\[-3mm] &&&& $1.1\times10^{18}$/cm$^{2}$/s \\
Jiao (2017)                 & $\sim$0.5 & 38 fs   & $1.6\times10^{5}$ & Isotropic with forward peak\\[-3mm] &&&& $6.7\times10^{16}$/s \\
Qi (2019)                   & 120       & 1.2 ps  & $3.2\times10^{6}$ & Isotropic\\[-3mm] &&&& $1.2\times10^{16}$ n/cm$^{2}$ \\
Feng (2020)                 & 3         & 30 fs   & $2.6\times10^{6}$ & Isotropic \\
G\"{u}nther (2022)           & 20        & 0.75 ps & $4.9\times10^{9}$ & Isotropic\\[-3mm] &&&& $10^{22}$ n/cm$^{2}$/s \\
Li (2022)                   & 3         & 45 fs   & $2.4\times10^{4}$ & Isotropic \\
Arikawa (2023)              & 4         & 30 fs   & $1.4\times10^{6}$ & Isotropic \\
Valli\`eres (2025)          & 3.2       & 22 fs   & $3.1\times10^{7}$ & Quasi-isotropic\\[-3mm] &&&& $1.0\times10^{17}$/cm$^{2}$/s \\
\hline
\end{tabular}
\caption{Laser wakefield accelerator photonuclear method neutron sources reported yields.   Kitagawa (2011) mechanism classification uncertain; verify whether electron-driven or TNSA.  Pomerantz (2014) is electron-driven from a solid density plasma, rather than LWFA}
\label{tab:lwfa_neutron_literature}
\end{table}

\section{Pitcher-catcher neutron yield scaling}
To determine the dependence of the neutron source properties on the laser inputs, we need the ion energy and total ion number accelerated by a short-pulse laser of normalized vector potential $a_0 = [2I/(n_c m_e c^3)]^{1/2}$ and total energy $\mathcal{E}_\ell\simeq Ic\tau_\ell A$ incident on a solid target.
The absorbed intensity $I_\mathrm{abs} = \eta I_\mathrm{laser} \propto \eta\, a_0^2$ drives hot electrons into the target.  Balancing the absorbed energy flux at the critical surface against the energy carried by the hot electron population streaming at $\sim c$ gives $\eta I_\mathrm{laser} = n_e^\mathrm{hot}\, T_e^\mathrm{hot}\, c$, so that $n_e^\mathrm{hot} = \eta I_\mathrm{laser}/(T_e^\mathrm{hot}\, c) \propto a_0^2/T_e^\mathrm{hot}$.
Integrating over the focal area $A$ and pulse duration $\tau_\ell$, the total number of hot electrons is $N_e^\mathrm{hot}\simeq n_e^\mathrm{hot}c\tau_\ell A=\eta \mathcal{E}_L/T_e^\mathrm{hot}$. Utilizing the TNSA sheath-field energy balance at the target rear surface, we estimate the number of ions in the beam as
\begin{equation}\label{eq:TNSAionnumberest}
  N_b \;\propto\; \frac{\eta\,\mathcal{E}_L}{T_e^\mathrm{hot}} \;\propto\; \frac{a_0^2}{T_e^\mathrm{hot}}\,,
\end{equation}
the final proportionality holding at fixed pulse energy.  

The rear-surface sheath electric field $|\vec E_s| \propto \sqrt{n_e^\mathrm{hot}\, T_e^\mathrm{hot}}$ accelerates ions.  An isothermal-expansion model predicts the maximum ion energy $E_{i,\mathrm{max}} \approx 2T_e^\mathrm{hot}[\ln(t_p + \sqrt{t_p^2+1})]^2$ with $t_p = \omega_{pi}\,t_\mathrm{acc}/\sqrt{2\exp(1)}$ and $t_\mathrm{acc}$ a characteristics acceleration timescale e.g. $\tau_\ell + t_\mathrm{hydro}$ \cite{Mor03}.
Since $|\vec E_s|^2 \propto n_e^\mathrm{hot}\, T_e^\mathrm{hot} \propto a_0^2$ (the density--temperature tradeoff cancels), $\varepsilon_\mathrm{max} \propto a_0^2$ independent of $T_e^\mathrm{hot}$ in the ultrashort-pulse limit, as observed experimentally.
However the slope of the beam spectrum depends on $T_e^\mathrm{hot}$: the Maxwellian slope parameter $T_p \sim T_e^\mathrm{hot}$ sets the fraction of ions above a given reaction threshold, so that the yield integral is exponentially sensitive to $T_e^\mathrm{hot}$ even though $E_{i,\mathrm{max}}$ is not.
For the hot electron temperature, Kluge~\textit{et~al.}\ show that proper-time averaging of the electron ensemble energy at the critical surface gives a scaling significantly below the standard ponderomotive result \cite{kluge2011electron}.  Specifically, their Eq.~(11), derived from the Riccati equation governing the transverse momentum in the evanescent field region, yields $T_e^\mathrm{hot}(a_0)$ in terms of an elliptic integral of the first kind $K(z)$,
\begin{align}\label{eq:Tescalingwa0}
  \frac{T_e^\mathrm{hot}}{m_ec^2} &=\frac{\pi}{2K(-a_0^2)}-1 \\
  &\approx\; \frac{\pi\, a_0}{2 \ln (16a_0)} \qquad (a_0 \gg 1)\,,
\end{align}
with the asymptotic behavior in the high intensity limit scaling as $\sim a_0/\ln a_0$ rather than $\sim a_0$.  The improved scaling at large $a_0$ has a significant impact on the yield estimate compared to the often-used Wilks scaling $T_e^\mathrm{hot}\sim a_0m_e$ \cite{Wilks92}. 

The average and maximum ion energy and total ion number control the beam-target nuclear reaction yield, which can be accurately computed in the thick-target limit \cite{Lab23}.  For a beam with Maxwellian spectrum $dN/dE_b \propto e^{-E_b/T_p}$ incident on a converter of density $n_t$ and length $L$, the yield of product $A$ is
\begin{align}\label{eq:beamfusionyield}
  Y_A &= N_b \int_0^\infty f(E_b;\,T_p)\,I_A(E_b)\,dE_b \,,\\
  I_A(E_0) &= \int_0^{E_0}\!\left(\frac{dE}{dx}\right)^{-1}\!\sigma_A\!\!\left(\frac{m_r}{m_b}E'\right)dE'\,,
\end{align}
where $I_A$ is the thick-target yield encoding the competition between the nuclear cross section $\sigma_A$ and the stopping power $dE/dx$.

Assembling Eqs. \eqss{TNSAionnumberest}{Tescalingwa0}{beamfusionyield} and noting that $T_p \propto T_e^\mathrm{hot}$, the yield as a function of $a_0$ is
\begin{equation}
\;Y_A(a_0) \;=\; \frac{\eta\,\mathcal{E}_L}{T_e^\mathrm{hot}(a_0)}\;\int_0^\infty f\!\bigl(E;\,T_p(a_0)\bigr)\;I_A(E)\;dE\;\;
\end{equation}
In the high-intensity limit, substituting Eq.~(2) into Eq.~(1) gives $N_b \propto a_0\ln a_0$.
The yield integral is controlled by the Gamow-like overlap between the exponential beam tail $\sim e^{-E/T_p}$ and the rising thick-target yield function $I_A(E)$ above the reaction threshold $E_\mathrm{th}$.  For a Maxwellian beam with $T_p \propto a_0/\ln a_0$, the dominant contribution to the integral comes from energies $E \sim E_\mathrm{th}$, giving a threshold suppression factor $g \sim \exp(-E_\mathrm{th}/T_p) = \exp(-E_\mathrm{th}\ln a_0/a_0)$.  Combining these, the yield scales with intensity as
\begin{equation}
Y_A(a_0) \propto a_0\,\ln a_0 \times g\!\left(\frac{E_\mathrm{th}\ln a_0}{a_0}\right)\;,
\end{equation}
where $g(x) \to \mathrm{const}$ for $x \ll 1$ (reactions well above threshold, $T_p \gg E_\mathrm{th}$) and $g(x) \sim e^{-x}$ for $x \gtrsim 1$ (reactions near or below threshold).
Thus at sufficiently high $a_0$ the yield grows as $a_0\ln a_0$, while near the onset of a reaction channel the exponential suppression dominates, producing a sharp turn-on whose location in $a_0$ is set by the condition $a_0/\ln a_0 \sim E_\mathrm{th}$.
The ponderomotive scaling would overestimate $T_e^\mathrm{hot}$ (and hence $T_p$) while underestimating $N_b$; these errors partially compensate in $\varepsilon_\mathrm{max}$ but not in the yield, because the thick-target yield integral weights the spectral shape nonlinearly.  This shows how the relativistically correct derivation of $T_e^\mathrm{hot}$ is essential for accurate yield predictions above $a_0 \sim 5$.

As a relevant example, we compare the impact of different $T_e^\mathrm{hot}$ scalings for deuteron break-up reactions ${}^{9}\mathrm{Be}(d,n)X$, ${}^{27}\mathrm{Al}(d,n)X$, and ${}^{63}\mathrm{Cu}(d,n)X$, common choices for converters or present as background apparatus.
Thick-target yield data (Labun 2022, building on M\'{e}nard \textit{et~al.}\ PRSTAB 1999 and Bem \textit{et~al.}\ PRC 2009) show that the ${}^{9}\mathrm{Be}(d,n)X$ thick-target yield $I_n(E)$ rises steeply from threshold and reaches $\sim 10^{-2}$ neutrons per deuteron by $E_d \approx 10$~MeV, plateauing near $\sim 10^{-1}$ at higher energies as the cross section (peaking near $\sigma \sim 800$~mb around 10~MeV) exits the range of available data.
The ${}^{27}\mathrm{Al}(d,n)X$ channels lie 1--2 orders of magnitude below, with per-deuteron yields of $\sim 10^{-3}$--$10^{-2}$ depending on whether only the ground-state ${}^{27}\mathrm{Al}(d,n){}^{28}\mathrm{Si}$ or the inclusive break-up cross section is used; ${}^{63}\mathrm{Cu}(d,n)X$ falls between the two and provides additional data constraining the $A$-dependence of the break-up cross section, which scales roughly as $A^{1/3}$ at fixed energy.
Crucially, the per-deuteron neutron probability measured at specific beam conditions---$w_\mathrm{Be} \approx 1.3 \times 10^{-3}$ and $w_\mathrm{Al} \approx 6.5 \times 10^{-5}$ from cross section calculations, compared with GEANT4 values of $w_\mathrm{Be} \approx 0.9 \times 10^{-2}$ and $w_\mathrm{Al} \approx 3.5 \times 10^{-2}$---already shows order-of-magnitude sensitivity to the treatment of the nuclear physics.
When folded over a Maxwellian beam distribution, the total yield $Y_i(T_p)$ exhibits the characteristic steep turn-on as $T_p$ sweeps through the effective threshold region: for ${}^{9}\mathrm{Be}(d,n)X$ the yield rises from $\sim 10^{-5}$ at $T_p \approx 1$~MeV to $\sim 10^{-2}$ at $T_p \approx 10$~MeV, while the ${}^{27}\mathrm{Al}$ channels track 1--2 orders of magnitude lower across the same range.
 
The Kluge--Wilks discrepancy in $T_p$ maps directly onto this steep portion of the yield curve.
At $a_0 = 20$ (roughly $10^{21}~\mathrm{W/cm}^2$), the Wilks scaling predicts $T_p \sim 6.7$~MeV while the Kluge scaling gives $T_p \sim 2.9$~MeV.
For ${}^{9}\mathrm{Be}(d,n)X$, reading off the yield-vs-$T_p$ curve, the Wilks prediction falls in the plateau region ($Y_i \sim 10^{-2}$) while the Kluge prediction sits on the rising slope ($Y_i \sim 10^{-3}$--$10^{-4}$), giving an order-of-magnitude discrepancy in the predicted neutron yield per deuteron---\emph{before} accounting for the additional factor of $\sim 2$ difference in $N_b$ between the two models.
For the lower-cross-section ${}^{27}\mathrm{Al}$ channels, which are still rising steeply at $T_p \sim 5$~MeV, the discrepancy is even more severe: the Wilks $T_p$ of 6.7~MeV places the beam well into the productive region, while the Kluge $T_p$ of 2.9~MeV leaves most deuterons below the effective threshold, producing yield predictions that can differ by a factor of $\sim 30$--$100$.
 
However, the errors from the two scalings again partially compensate in the \emph{total} neutron number $Y_n = N_b \times Y_i(T_p)$: the Wilks scaling underpredicts $N_b$ (by $\ln a_0/\sqrt{2} \approx 2$) but overpredicts $Y_i$ (by $\sim 10$--$100$), so the net Wilks prediction for total neutron yield substantially \emph{overpredicts} the Kluge result.
This partial compensation does \emph{not} rescue the Wilks prediction---the exponential sensitivity of $Y_i$ to $T_p$ in the threshold region overwhelms the linear correction in $N_b$.
As with the proton-boron case, yield \emph{ratios} between different converter materials provide a clean discriminant.
Since the ${}^{9}\mathrm{Be}$ and ${}^{27}\mathrm{Al}$ thick-target yields have different effective thresholds and different slopes in the turn-on region, the ratio $Y_n^\mathrm{Be}/Y_n^\mathrm{Al}$ is a strong function of $T_p$ in the 2--8~MeV range.
The $N_b$ prefactor and all geometric factors cancel in this ratio, leaving a quantity that depends only on the beam spectral shape---and hence on the electron temperature scaling through $T_p(T_e^\mathrm{hot}(a_0))$.
Experimentally, a laser shot that irradiates both a Be and an Al converter (or sequential shots at the same $a_0$) and measures the neutron yield from each would determine $T_p$ from the yield ratio, which can then be checked against the Kluge prediction $T_p \propto a_0/\ln a_0$ versus the Wilks prediction $T_p \propto a_0$.
The data from the Texas Petawatt (shots 11557 and 11555, giving $N_d \sim 10^{12}$--$10^{13}$ deuterons and $N_n \lesssim 5 \times 10^{9}$ neutrons from Be) already constrain the product $N_b \times Y_i(T_p)$; combined with the measured beam spectrum from RCF, they provide an initial benchmark.
More precise measurements of the Be-to-Al neutron yield ratio at varying $a_0$ would trace out the yield-ratio curve and provide a stringent test of which temperature scaling correctly describes the front-surface interaction, entirely independent of electron spectroscopy.
\end{appendix}

\end{document}